\RequirePackage{fix-cm}
\documentclass{svjour3}                     
\smartqed  
\usepackage{graphicx}
\usepackage{color}
\usepackage{mathptmx}      
\usepackage{amsmath,amsfonts}
\usepackage[numbers,sort&compress]{natbib}

\DeclareMathOperator*{\esssup}{ess\,sup}

\DeclareMathOperator*{\argmax}{arg\,max}

\renewcommand{\P}{\mathbb{P}}

\newcommand{\R}{\mathbb{R}}
\newcommand{\E}{\mathbb{E}}

\newcommand{\N}{\mathbb{N}}

\newcommand{\cA}{\mathcal{A}}
\newcommand{\cF}{\mathcal{F}}

\newcommand{\cI}{\mathcal{I}}

\newcommand{\eps}{\varepsilon}

\allowdisplaybreaks[4]

\DeclareMathOperator\sgn{sgn}

\newcommand{\nada}[1]{}

\begin{document}

\title{Teamwise Mean Field Competitions
}


\author{Xiang Yu
        \and
        Yuchong Zhang
        \and
        Zhou Zhou 
}
\authorrunning{X. Yu et al.} 

\institute{Xiang Yu \at
              Department of Applied Mathematics, The Hong Kong Polytechnic University, Kowloon, Hong Kong. \\
              \email{xiang.yu@polyu.edu.hk}           
           \and
           Yuchong Zhang \at
              Department of Statistical Sciences, University of Toronto, 100 St. George Street, Toronto, Canada. \\
              \email{yuchong.zhang@utoronto.ca}
           \and
           Zhou Zhou \at
              School of Mathematics and Statistics, The University of Sydney, NSW 2006, Sydney, Australia. \\
              \email{zhou.zhou@sydney.edu.au}
}

\date{Received: date / Accepted: date}

\maketitle

\begin{abstract}
This paper studies competitions with rank-based reward among a large number of teams. Within each sizable team, we consider a mean-field contribution game in which each team member contributes to the jump intensity of a common Poisson project process; across all teams, a mean field competition game is formulated on the rank of the completion time, namely the jump time of Poisson project process, and the reward to each team is paid based on its ranking. On the layer of teamwise competition game, three optimization problems are introduced when the team size is determined by: (i) the team manager; (ii) the central planner; (iii) the team members' voting as partnership. We propose a relative performance criteria for each team member to share the team's reward and formulate some special cases of mean field games of mean field games, which are new to the literature.
In all problems with homogeneous parameters, the equilibrium control of each worker and the equilibrium or optimal team size can be computed in an explicit manner, allowing us to analytically examine the impacts of some model parameters and discuss their economic implications. Two numerical examples are also presented to illustrate the parameter dependence and comparison between different team size decision making.
\end{abstract}

\section{Introduction}
We are interested in some unconventional mean field games in which each ``agent" is itself a team of many, many individuals having their own interactions. Mathematically speaking, some special cases of mean field games of mean field games, or two-layer mean field games, are formulated and investigated for the first time. In particular, we consider the setting that the top layer of inter-team problem is a rank-based reward competition game, and the bottom layer of intra-team problem is a contribution game in which team members collaborate towards the completion of a common, reward-earning project.

Our research is mainly motivated by ubiquitous real life competitions of large firms aiming for new product or technology development, such as the recent worldwide competition for 5G network technology or some mega-institutions preempting their opponents in the invention of new medical devices. Unlike the mean field competition games based on single agents as in \cite{Bay-Zhang.16, Nutz-Zhang.19, BCZ.19, Bay-Zhang.19}, teamwork formulation plays a core role in the present paper, which is rarely discussed in the existing literature, but certainly deserves a more careful investigation. In fact, the majority of real life competitions occur on a group basis. In industries such as manufacturing, service and technology development, organizing workers into teams has been shown to be an effective way to enhance productivity and teamwork can undoubtedly accelerate the completion time of the desired project comparing to a single agent. On the other hand, a large team size may also have adverse impacts towards its total profit. Take the company organization as an example, labor costs and other size-related costs such as office renting, insurance payment, human resource management fees will directly prevent the unlimited growth of the company size. More importantly, as the company size gets larger, it has been observed in real life that employees are more likely to become free-riders or exhibit gradualism and share the team's reward without making much effort, see related research in \cite{legros1993efficient, lockwood2002gradualism, compte2004gradualism, yildirim2006getting, bonatti2011collaborating, battaglini2014dynamic, campbell2014delay} and the references therein. Thus, when teams, especially large teams, are taken into account, several questions naturally arise regarding each employee's working efficiency and organizational issues: How should members of a team share the reward more efficiently, how will the team size affect the members' incentive and the company's total profit, and by what metric can we define and quantitatively choose the optimal team size, and if teams interact with each other, the equilibrium team size?

Dynamic contribution game in a single team with finite agents has been studied in \cite{Georgiadis.15}, where the project process is driven by a Brownian motion, and members jointly contribute to its drift (see also \cite{CviGeo.16} for an efficient mechanism design in the absence of noise).
\cite{Georgiadis.15} insightfully addressed the question of efficient team size under two types of reward sharing schemes: \textit{the public good allocation scheme}, wherein each agent receives a reward irrespective of the team size; and \textit{the budget allocation scheme}, wherein the agents share the total reward paid directly to the team. Moreover, size impacts are examined in two scenarios: \emph{corporation with a team manager} and \emph{self-organized partnership}. In the first case, the manager recruits a group of agents working for him and determines the team size to maximize his own expected payoff, while in the second case, the project's reward is directly shared among all team members and the team size is decided by public voting.

Inspired by \cite{Georgiadis.15}, we also seek to consider the two aforementioned reward allocation schemes and the teamwork formulation of top layer corporation and partnership in the present paper. On top of that, teams do not live in isolation, but are part of the completion time ranking game similar to \cite{Nutz-Zhang.19}. As opposed to the competition game of single agents as in \cite{Nutz-Zhang.19} or the contribution game of workers in a single team as in \cite{Georgiadis.15}, we are interested in the teamwise competition game, which involves a sophisticated two-layer interaction. The mutual dependence and inherently coupled decision making, especially when the number of teams and team members are finite, may render the equilibrium individual control impossible to obtain. As a remedy, we formulate the teamwise competition game in the mean field sense so that the problem enjoys the decentralized structure and the impact of an individual worker or a team on the aggregate distribution is negligible. We can thus formally solve a stochastic control problem for a representative worker against a fixed environment together with some consistency conditions. To make the two-layer mean field game tractable, we adopt the simplest dynamics for our project process: an inhomogeneous Poisson process with controlled intensity that equals the aggregate effort of all team members. In addition, we assume homogeneity both at the team level and at the team member level, while leaving the extension to the heterogeneous case for future study. In a teamwise competition, how the team's reward are shared by its members becomes crucial. We not only plan to formulate a payoff for each worker such that the fixed point argument for the two-layer mean field game can be carried out, but also need to design the incentive scheme in a reasonable way to prevent free-riders or gradualism when the team size is very large. One of our main contributions is to propose a relative performance criterion, see the objective function in \eqref{eq:Vz} with a bonus reward, such that the equilibrium control can be derived explicitly, which also carries an essential encouragement effect that can successfully eliminate the unpleasant free-riding behavior. Furthermore, in both the corporation model with team manager (see Theorem \ref{thm:manager}) and the partnership model with self-organization (see Theorem \ref{thmpartner}), the top layer equilibrium team size can be explicitly characterized based on model parameters. In particular, we identify conditions on the parameters when the equilibrium is not to assemble the team, i.e., when there is no sufficient incentive for the team manager or the team member to join in the teamwork for competitions. Some quantitative impacts of the model parameters, such as how much bonus amount based on relative performance to pay the workers and how much wage to pay the team manager, on a representative worker's value function and the equilibrium team size can be analyzed, which are consistent with observations in real life organizations.

On the top layer, a third team size decision making problem is also considered that hinges on our tractable teamwork formulation with homogeneous parameters: a central society planner, such as the government, assigns a unified size for all teams by maximizing the average team member's welfare (see Theorem \ref{thm:socialplan}). In particular, one can interpret this formulation as a toy model of planned economy when the market consists of many large state-owned enterprises and the government decides the firm size for all. The solvable centralized optimization problem leads to possible comparison with the model where the team size is determined by the team manager under market competitions. The latter case can be understood as a toy model of market economy. Along this line, we are able to numerically illustrate some interesting observations, from the perspective of mean field competitions, that the market economy and planned economy have both pros and cons depending sensitively on the market environment and different parameters.

Recently, there are some emerging work on mean field games and mean field control with multiple populations, see \cite{Benseveral} and \cite{Fujii} for instance, which are relatively close to the feature of our two layer mean field game. Nevertheless, our work differs substantially from \cite{Benseveral} and \cite{Fujii} in that the number of populations or teams is infinity and the team size is also part of the control. Furthermore, our objective functional is defined as the expected rank-based reward minus effort and size costs while \cite{Benseveral} and \cite{Fujii} mainly focus on the standard linear quadratic payoff. On the other hand, some recent work such as \cite{7799341}, \cite{dylan-1} and \cite{SANJ-1} studied the mean field team problem for a single team with mean field interacting members, but the teamwise interaction and the equilibrium team size choice are not considered therein. The formulation of a mean field game of mean field games makes the present paper distinct from the existing literature and mathematically interesting. Apart from heterogeneous teams, potential future extensions also include diffusion framework and incorporating dynamic team size control instead of the current static team size decision making.

The rest of the paper is organized as follows. Section \ref{sec-2} introduces the teamwork formulation and the bottom layer of intra-team mean field contribution game. The relative performance payoff is proposed and the equilibrium control of a representative worker is solved using dynamic programming arguments. Section \ref{sec-3} further considers the top layer of inter-team competition and the team size decision making. Three distinctive models are studied and the explicit equilibrium or optimal team size is derived in each problem. In Section \ref{sec-4}, based on the explicit results in the previous models, numerical examples are provided to illustrate some quantitative impacts of model parameters on the equilibrium team size and the value of a representative worker. Some comparison analysis between different team size decision making are also performed. Proofs of main results are given in the Appendix \ref{sec:proof}.

\ \\
\ \\
\textit{List of model parameters.}\\
For the readers' convenience, we summarize here a list of all model parameters and their meanings.\\
$-K$: the total size of the rank-based reward, excluding division effect;\\
$-p$: convexity parameter of the rank-based reward; \\
$-\eps$: intra-team devision effect;\\
$-\beta$: proportion of the fixed salary part for a representative worker;\\
$-\theta$: proportion of a team's reward allocated to its manager;\\
$-c$: cost of effort parameter for a representative worker; \\
$-\kappa_0$: fixed cost to build a team;\\
$-k$: coefficient of the variable size cost to build a team;\\
$-\delta$: convexity parameter of the variable size cost to build a team.

\section{Mean Field Collaboration Within A Team}\label{sec-2}

\subsection{Probabilistic Setup}\label{sec:prob-setup}
We adopt the probabilistic setup of \cite{Nutz-Zhang.19}. Let $(I, \cI, \mu)$ be an atomless probability space, in which each $i\in I$ represents a team. Let $(\Omega, \cF, \P)$ be another probability space supporting a family $(Z^i)_{i\in I}$ of exponential random variables with the unit rate. We assume that the family $(Z^i)_{i\in I}$ is essentially pairwise independent, meaning for $\mu$-almost all $i\in I$, $Z^i$ is independent of $Z^j$ for $\mu$-almost all $j\in I$. In addition, it is assumed that $(Z^i)_{i\in I}$ is defined on an extension of $(I\otimes \Omega, \cI\otimes \cF, \mu\otimes \P)$ for which the Exact Law of Large Numbers (see \cite{Sun.06}) holds.

Define the set $\cA_0$ of admissible feedback controls as the set of functions $\lambda:[0,1]\rightarrow \R_+$ that are locally piecewise Lipschitz continuous on $[0,1)$.
For any $\lambda\in \cA_0$, there exists a unique continuous solution $\rho_\lambda:\R_+\rightarrow [0,1]$ satisfying
\begin{equation}\label{eq:rho}
\rho(t)=\int_0^t \lambda(\rho(s))(1-\rho(s))ds, \quad t\ge 0.
\end{equation}
Note that $\rho\in [0,1)$ if $\lambda$ is bounded.
We further define the stopping time
\[\tau^i_\lambda:=\inf\left\{t\ge 0: \int_0^t \lambda(\rho_\lambda(s))ds=Z^i\right\}, \quad i\in I.\]
The family $(\tau^i_\lambda)_{i\in I}$ is essentially pairwise independent and corresponds to the family of first jump times of independent inhomogeneous Poisson processes with intensity $\lambda\circ\rho_\lambda$.
By the Exact Law of Large Numbers, we have the following lemma (cf.\ \cite[Lemma 1]{Nutz-Zhang.19}).
\begin{lemma}
Suppose all teams employ the jump intensity $\lambda \in \cA_0$, then
\begin{align*}
\rho_\lambda(t)&=\mu\left\{i: \tau^i_\lambda(\omega)\in [0,t]\right\} \quad \P\mbox{-a.s.}\\
&=\P\left(\tau^i_\lambda \in [0,t]\right) \quad \mu\mbox{-a.s.}
\end{align*}
In other words, $\rho_\lambda(t)$ is both the proportion of teams that have jumped (completed their projects) by time $t$ and the probability that a representative team has jumped (completed its project) by time $t$.
\end{lemma}

We will restrict ourselves to controls belonging to a subset of $\cA_0$, defined by
\[\cA:=\left\{\lambda\in \cA_0:  \lambda(r)>0,  \; \forall\, r\in [0,1) \quad \text{and}\quad  \int_0^{1}\frac{(1-x)^{2p-1}}{\lambda(x)}dx<\infty \right\},\]
where $p$ is a positive parameter to be introduced in Section~\ref{subsec:payoff}.

\subsection{Relative performance criterion with encouragement effect}\label{subsec:payoff}

Let us write $\rho(t):=\rho_\lambda(t)$ for simplicity and consider a representative team (say, team $i$) with a continuum of players, defined on an atomless finite non-negative measure space $(J, \mathcal{J}, \nu)$.  We refer to $z:=\nu(J)$ as the size of the team and it is assumed to be fixed in the rest of this section.

We assume that each team member or worker observes $(\rho(t))_{t\geq 0}$, and this assumption is imposed throughout the paper.\footnote{Using the information generated by $\rho(t)$ would lead to the closed loop equilibrium in the two-layer mean-field games that will be investigate later. As is well known in the literature of mean-field game, such closed loop equilibrium would also give an open loop equilibrium, which means that each team member does not need to observe $\rho(t)$ at the equilibrium.} Consequently, each team member chooses her effort rate $\alpha\in \cA$ as a function of $\rho(t)$, which then gets aggregated into the jump intensity of the team's Poisson project process. Suppose that all other workers in this team choose the feedback control $\bar\alpha\in \cA$, the team's project process will follow an inhomogeneous Poisson process with jump intensity  $z\bar\alpha\circ \rho$.\footnote{One can also think of the team sizes as determining the change of measure from $\mathbb{P}$ to some $\mathbb{Q}$ under which $Z^i$ is an exponential random variable with rate $z_i$.} Let us denote by $\tau:=\tau^i_{z\bar\alpha}$ its first jump time which signals the completion of the project by this team.

A representative worker in the team needs to solve the stochastic control problem in a fixed environment defined by
\begin{equation}\label{eq:Vz}
\sup_{\alpha\in\mathcal{A}}\E\left[G_z(\rho(\tau))\left(\beta+(1-\beta)\frac{\alpha(\rho(\tau))}{\bar\alpha(\rho(\tau))}\right)-c\int_0^\tau\alpha(\rho(t))^2\,dt\right],
\end{equation}
where $c>0$ is the cost of effort parameter, $G_z(r)$ is the per person reward paid to a team of rank $r\in [0,1]$, $\beta\in [0,1)$ is a proportion parameter to be introduced, and $\rho(\tau)$ is the rank of team $i$, measured as the proportion of teams that have jumped before or at the same time as team $i$ (so that $r=0$ corresponds to the top rank and $r=1$ corresponds to the bottom rank). We use the convention that $\rho(\infty):=1$ and $\alpha(1)/\bar\alpha(1):=1$, meaning that teams who never complete the project receive the bottom rank, and all members of bottom ranking teams receive $G_z(1)$ regardless of their relative performance within the team.

It is assumed henceforth in this paper that the reward function $G_z$ takes the specific form:
$$G_z(r)=K(1+p)(1-r)^p\cdot z^{-\eps}, \quad K, p>0,\, \eps\in [0,1].$$
Here $K$ represents the total reward pie shared by all teams, and $p$ depicts how skewed the reward is towards top ranking teams. A large $p$ value indicates that most of the reward is given to highly ranked teams, which is consistent with the wealth distribution of our society that some top ranked groups earn the majority chunk of the social wealth pie. Moreover, $\eps$ here captures the intra-team division effect due to the team size. For example, if $\eps=0$, there is no division effect and $G_z$ corresponds to the \textit{public good allocation scheme} under which each agent receives the same reward regardless of the team size. If $\eps=1$, the total reward for the team is relatively fixed and $G_z$ corresponds to \textit{the budget allocation scheme} as all (unit measure of) team members will share a fixed pie.

Our proposed payoff in \eqref{eq:Vz} is said to be a relative performance criterion. This is reflected both at the team's level via rank-based rewards and at the team member's level via the term $\alpha/{\bar\alpha}$, which captures the relative effort when the milestone of the project is achieved.
It is worth noting that the payoff consists of a fixed salary part $\beta G_z(\rho(\tau))$ as well as a performance-based bonus part $(1-\beta) G_z(\rho(\tau))  \frac{\alpha}{\bar\alpha} (\rho(\tau))$, where the proportion $\beta \in [0,1)$ determines how the rank-based reward is split between the two parts.\footnote{To simplify the presentation, let us assume that $\beta$ is a constant. The main results can be easily extend to the cases when $\beta$ is a function of the team size $z$. In particular, Theorem~\ref{thm:EQMeffort} remains valid.} In principle, the bonus term from relative effort can stimulate the diligence of team members.

In some real life situations, the bonus income may dominate the regular salary part significantly. Admittedly, it may not be the best choice to pay the bonus based on the effort at the completion time directly. However, as the completion time is characterized as the jump time of a Poisson process which is memoryless, each worker has no incentives to wait and be lazy, because she may miss the chance to make higher efforts when the jump time suddenly occurs. This incentivizes everyone to be diligent from the beginning and compete with peers by sustaining higher efforts than others at the completion time. This feature in our formulation can be interpreted as the encouragement effect that may prevent free-riding in large teams. Mathematically speaking, our choice of the control ratio $\alpha(\rho(\tau))/\bar{\alpha}(\rho(\tau))$ at the completion time in the bonus income term remarkably yields fully explicit value function and best response control of the representative worker. These explicit formulas significantly simplify the fixed point argument for the inter-team equilibrium, rendering some economic observations and implications possible.

\subsection{Equilibrium control for each team member}
Having introduced our objective functional for the representative worker, we proceed to study her best response control when all her co-workers use control $\bar{\alpha}\in \cA$. Throughout this subsection, the size $z$ of the representative team and the jump intensity $\lambda\in\cA$ (hence $\rho$) of all other teams are given and fixed. 

As the dynamic version of \eqref{eq:Vz}, the value function before the completion of the project is defined as
\[V(r)=V(r;\lambda,z,\bar\alpha):=\sup_{\alpha\in\mathcal{A}} J(r,\alpha ;\lambda, z, \bar\alpha),\quad r\in[0,1],\]
where
\begin{align*}
J(r,\alpha ;\lambda, z,\bar\alpha):=\E\left[G_z(\rho(\tau)) \cdot\left(\beta+(1-\beta)\frac{\alpha(\rho(\tau))}{\bar\alpha(\rho(\tau))}\right)-c\int_0^\tau\alpha^2(\rho(t))\,dt\,\Big| \rho(0)=r\right],
\end{align*}
and $\rho$ satisfies \eqref{eq:rho}. Assume $\lambda\in\mathcal{A}$. The associated dynamic programming equation can be heuristically derived as
\begin{equation}\label{DPE}
\lambda(1-r) V'+\sup_{\alpha\ge 0}\left\{\left[G_z \cdot \left(\beta+(1-\beta)\frac{\alpha}{\bar\alpha}\right)-V\right] \cdot z\bar\alpha -c\alpha^2\right\}=0,
\end{equation}
with the boundary condition $V(1)=0$.

By the first order condition, a candidate best response control is given by
\begin{equation}\label{alphaz}
\alpha_{z}(r)=\frac{(1-\beta)zG_z(r)}{2c}.
\end{equation}
Note that $\alpha_{z}(r)\equiv 0$ if $\beta=1$, i.e. if there is no bonus income. This is what we called the free-riding behavior as the best response is to do nothing and simply wait for the fixed salary generated by co-workers, due to the lack of incentives for diligence. This justifies the necessity for us to introduce the bonus income based on relative effort in the teamwork formulation with a large population base.

We say $\alpha^*\in\mathcal{A}$ is an intra-team equilibrium control of the representative worker if it is the best response to itself. This imposes the consistency condition on the best response control that
$$\bar\alpha=\alpha_{z}=\frac{(1-\beta)zG_z}{2c}.$$
Substituting the above $\bar{\alpha}$ and $\alpha_{z}$ back into \eqref{DPE} yields
\begin{equation}\label{DPE2}
\lambda(1-r) V'-\frac{1-\beta}{2c}z^2G_z\cdot V+\frac{1-\beta^2}{4c}z^2G_z^2=0,\quad V(1)=0.
\end{equation}
This is a first order linear ODE that can be solved explicitly, which leads to the following main result of this section. A rigorous proof is postponed to Section \ref{sec:proof}.
\begin{theorem}\label{thm:EQMeffort}
Let the size $z>0$ of a team and the jump intensity $\lambda\in\cA$ of all other teams be given.
There exists a unique equilibrium effort $\alpha_{z}\in\mathcal{A}$ within the team, given by \eqref{alphaz}.
In equilibrium, the value function of each team member before the completion of the project, denoted by $V_{\lambda,z}$, is 
{\small
\begin{align}
\notag V_{\lambda,z}(r)&=\frac{1-\beta^2}{4c}{z^2}\int_r^1\frac{G_z^2(x)}{\lambda(x)(1-x)}\cdot\exp\left(-\frac{1-\beta}{2c}z^2\int_r^x\frac{G_z(y)}{\lambda(y)(1-y)}dy\right)\,dx\\
\label{DPE2soln} &=\frac{1-\beta^2}{4c}K^2 (1+p)^2 z^{2-2\eps}\int_r^1\frac{(1-x)^{2p-1}}{\lambda(x)}\cdot\exp\left(-\frac{K(1+p)(1-\beta)}{2c} z^{2-\eps}\int_r^x\frac{(1-y)^{p-1}}{\lambda(y)}dy\right)\,dx.
\end{align}
}
\end{theorem}

\begin{remark}
From \eqref{DPE2soln}, we can see that if $\eps\in[0,1)$, then $V_{\lambda,z}(0)\to 0$ as $z\to 0$. Intuitively, if the size of the team is very small, it is very likely that the team's ranking will not be good due to insufficient labor. A more interesting observation is that $V_{\lambda,z}(0)\to 0$ as $z\to\infty$ even if $\eps=0$. An explanation is that, if the team size is very large, the unexpected jump may happen very soon. As the payoff for each worker depends on her relative contribution, she will try to maintain high effort at any time in order to get a higher reward when the jump happens, which results in a significant cost of effort that offsets the benefit of an increased reward. In other words, the larger the team size, the higher peer competition and peer pressure each worker may confront within the team in our setting with bonus income.
\end{remark}

\begin{remark}\label{rmk:zerolambda}
The above theorem is valid under the assumption that $\lambda$ is strictly positive on $[0,1)$. If $\lambda\equiv 0$, which may be a result of the other teams choosing zero team size (assuming it is allowed), then the state process $\rho\equiv \rho(0)$ and we have that for $z>0$, $r<1$,
\begin{align*}
J(r,\alpha ;0, z,\bar\alpha)&=G_z(r) \cdot\left(\beta+(1-\beta)\frac{\alpha(r)}{\bar\alpha(r)}\right) \P(\tau<\infty)-c\alpha^2(r)\E\tau\\
&=G_z(r) \cdot\left(\beta+(1-\beta)\frac{\alpha(r)}{\bar\alpha(r)}\right)-\frac{c\alpha^2(r) }{z\bar \alpha(r)}\\
&=\beta G_z(r)+\frac{1}{\bar \alpha(r)}\left[G_z(r)(1-\beta)\alpha(r)-\frac{c}{z}\alpha^2(r)\right],
\end{align*}
where we have used that the controlled jump time $\tau=\tau_{z\bar \alpha}$ is exponentially distributed with rate $z\bar \alpha(r)$. Here $r=\rho(0)$ is a fixed number, and $r>0$ means that an $r$-fraction of teams have already jumped and the remaining ones have zero team size and thus zero jump intensity. The unique intra-team equilibrium effort in this case is again given by \eqref{alphaz} with associated value
\begin{equation}\label{eqq2}
V_{0,z}(r)=\frac{1+\beta}{2}G_z(r).
\end{equation}
\end{remark}

The quantity $V_{z\alpha_z, z}(0)$ will play an important role in the later analysis. It represents the expected payoff of a team member if all of her peers, within or outside the team, use control $\alpha_z$, and all teams are of the size $z>0$. To ease future reference, we compute it from \eqref{DPE2soln} that
\begin{equation}\label{EQM-V0}
V_{z\alpha_z, z}(0)=\frac{K(1+\beta)}{2} z^{-\eps}.
\end{equation}
Recall that $K$ is the total rank-based reward shared by all teams (or average reward per team since the team space has measure one).
The factor $(1+\beta)/2< 1$ accounts for the loss due to the cost of effort.

\section{Mean Field Competition Among Teams}\label{sec-3}
This section begins to study the top layer game among a continuum of teams which are assumed to be homogeneous. The game starts fresh with $\rho(0)=0$, and all teams compete for the ranking of completion times which are modeled as the first jump time of independent inhomogeneous Poisson processes (see the model setup in Section~\ref{sec:prob-setup}). From this point onwards, the team size is no longer priorly fixed. Instead, it is the team's task to determine the team size for its own profit. In particular, we intend to derive the equilibrium solution of the two-layer mean field game and investigate the impacts of the team size on the representative worker's effort and expected payoff.

The benefit for having a larger team size is the greater likelihood of an earlier jump time. On the flip side, assembling a team is costly. To account for various size-related costs, we introduce a fixed cost $\kappa_0\ge 0$ to build the team as well as a non-decreasing variable size cost $\kappa(z)$ for each team, defined as
$$\kappa(z)=k\cdot z^\delta,\quad k, \delta>0.$$
In the subsequent sections, we will examine three different types of decision making over the team size and the equilibrium solution of the associated two-layer mean field games.

\subsection{Equilibrium Team Size by Team Managers}\label{subsec:m}
In this section, we assume that each team has its own manager, who decides the optimal team size $z$ at the initial time. The team manager recruits team members to work on his behalf, and he will get a proportion $\theta \in(0,1)$ of the team's total gain $ K(1+p)(1-\rho(\tau))^p =z^\eps G_z(\rho(\tau))$ and allocate the remaining reward to all team members.\footnote{Here the intra-team division effect only applies to the regular team members' share of the reward, namely, $(1-\theta)K(1+p)(1-\rho(\tau))^p$. In the public good allocation scheme ($\eps=0$), the manager and the each member's reward have the same order of magnitude; in the budget allocation scheme ($\eps=1$), the manager receives a chunk of the fixed pie while each member shares a negligible piece of the remaining pie.}

Meanwhile, the manager is responsible for all size-related costs for the team; that is,
he needs to pay
the fixed cost $\kappa_0$ to build the team and the variable size cost $\kappa(z)$ given that the team is successfully assembled with positive size $z$.
Along this line, given the team size $z>0$, each worker or regular team member faces the optimization problem:
\begin{align*}
&\sup_{\alpha\in \cA}\E\left[(1-\theta)G_z(\rho(\tau))\cdot\left(\beta+(1-\beta)\frac{\alpha(\rho(\tau))}{\bar\alpha(\rho(\tau))}\right)-c\int_0^\tau\alpha(\rho(t))^2\,dt\right]\\
&=(1-\theta)\sup_{\alpha\in \cA}\E\left[G_z(\rho(\tau))\cdot\left(\beta+(1-\beta)\frac{\alpha(\rho(\tau))}{\bar\alpha(\rho(\tau))}\right)-\tilde c\int_0^\tau\alpha(\rho(t))^2\,dt\right],
\end{align*}
where $\tilde c:=c/(1-\theta)$.

Let us focus on a representative team, again named $i$.
Suppose that all other teams pick a jump intensity $\lambda\in \cA$ or $\lambda\equiv 0$ and that the manager of team $i$ chooses a team size of $z> 0$. In view of Theorem~\ref{thm:EQMeffort} and Remark~\ref{rmk:zerolambda}, the intra-team equilibrium effort is
\begin{equation}\label{eq:alphaz2}
\alpha_{z}=\frac{(1-\beta)}{2\tilde c}zG_z,
\end{equation}
which lies in $\cA$ and happens to be independent of $\lambda$.
Having the equilibrium effort in mind, the manager of team $i$ then maximizes his own objective function
\begin{align*}
J^m(z;\lambda):=1_{\{z>0\}}\left(\theta \mathbb{E}\left[ z^\eps G_z(\rho_\lambda(\tau_{\lambda,z}))\right]-\kappa_0-\kappa(z)\right)
\end{align*}
over size $z\ge 0$, where $\tau_{\lambda,z}$ is the jump time of team $i$ when all of its members use control $\alpha_{z}$.
Here we allow the manager to choose zero team size, meaning that the team is not assembled and will never complete the project.
We set $\alpha_0:\equiv 0$
for the case $z=0$, which is consistent with the formula for positive $z$.

\begin{remark}
It is worth noting that our team manager model differs from the standard principal-agent problem with infinitely many agents (e.g.\ \cite{dylan-1} and \cite{CarWang}). The role of the team manager in our model is to statically decide the constant team size at the initial time instead of controlling the incentive of each agent. The reward and bonus income weight to each worker are fixed and not determined by the team manager. Moreover, all team managers interact with each other through the inter-team competition. It will be more interesting yet more complicated to formulate the intra-team problem with a team manager as a principal-agent problem in future research, in which the principal can control the reward paid to the workers as well as a team size that dynamically evolves over time.
\end{remark}

We now give the exact definition of a Nash equilibrium in the two-layer mean field game.

\begin{definition}
A pair $(z^\ast,\alpha^\ast)\in \R_+ \times \cA$ is said to be an equilibrium for the two-layer mean field game with manager-selected team sizes if
\[ z^* \in \argmax_{z\ge 0} J^m(z; z^\ast \alpha^\ast)\quad \text{and}\quad \alpha^\ast=\alpha_{z^\ast}.\]
That is, $z^\ast$ is the optimal size of a representative team given that all other teams choose size $z^*$ and individual control $\alpha^\ast$; and $\alpha^\ast$ is the intra-team equilibrium control given that all teams including the representative one have size $z^\ast$.
We refer to such a $z^\ast$ as an equilibrium team size and such an $\alpha^\ast$ as the associated equilibrium effort.
\end{definition}

By definition, $z^\ast$ is an equilibrium team size if and only if
either
\begin{align*}
z^\ast=0 \quad \text{and} \quad \sup_{z>0} \left\{\theta z^\eps G_z(0)-\kappa_0-\kappa(z)\right\} \le 0,
\end{align*}
or
\begin{align*}
z^\ast>0 \quad \text{and} \quad z^\ast & \in \argmax_{z\ge 0} \left\{1_{\{z>0\}}\left(\theta \mathbb{E}\left[ z^\eps G_z(\rho_{\lambda^\ast}(\tau_{\lambda^\ast,z}))\right]-\kappa_0-\kappa(z)\right)\right\},
\end{align*}
where
\begin{equation}\label{lambdastar}
\lambda^\ast=z^\ast \alpha_{z^\ast} =\frac{(1-\beta)}{2\tilde c}(z^\ast)^2 G_{z^\ast}.
\end{equation}
The first case occurs if and only if $K(1+p)\theta\le \kappa_0$. In the second case, the quantity $\E[G_z(\rho_{\lambda^\ast}(\tau_{\lambda^\ast, z}))]$ can be computed as the initial value of the solution $v(r)$ to the following terminal value problem:
$$\lambda^\ast(1-r) v'-\frac{1-\beta}{2\tilde c}z^2G_z\cdot v+\frac{1-\beta}{2\tilde c}z^2G_z^2=0,\quad v(1)=0.$$
Solving the above first order ODE, we obtain that
\begin{align*}
\E[G_z(\rho_{\lambda^*}(\tau_{\lambda^\ast,z}))]&=v(0)\\
&=\left(\frac{z}{z^\ast}\right)^2 \int_0^1 \frac{G_z^2(r)}{(1-r)G_{z^\ast}(r)} \exp\left(- \left(\frac{z}{z^\ast}\right)^2\int_0^r \frac{G_z(y)}{(1-y)G_{z^\ast}(y)}dy\right)dr\\
&=K(1+p) z^{-\eps} \left[1+p \left(\frac{z}{z^\ast}\right)^{\eps-2}\right]^{-1}.
\end{align*}
Hence, the set of positive equilibrium team sizes can be characterized as all points $z^\ast>0$ satisfying
\begin{equation}\label{eq:z*0}
z^\ast\in \argmax_{z> 0} \left\{K(1+p)\theta \left[1+  p \left(\frac{z}{z^\ast}\right)^{\eps-2} \right]^{-1}-\kappa(z)\right\}
\end{equation}
and
\begin{equation}\label{eq:zstar-b}
K\theta-\kappa_0-\kappa(z^\ast)\ge 0.
\end{equation}
The expression of $\E[G_z(\rho_{\lambda^*}(\tau_{\lambda^*,z}))]$ also implies that at any equilibrium $(z^\ast, \alpha^\ast)$ with $z^\ast>0$, the value $V^m$ for the team manager is given by
\begin{equation}\label{eq:Vm}
V^m=\theta\E[ (z^\ast)^{\eps} G_{z^\ast}(\rho_{\lambda^*}(\tau_{z^\ast}))]-\kappa_0-\kappa(z^\ast)=K\theta-\kappa_0-\kappa(z^\ast),
\end{equation}
which is guaranteed to be non-negative by \eqref{eq:zstar-b}.
Meanwhile, the value function $V^w$ of each regular team member or worker can be computed from \eqref{EQM-V0}, giving that for $z^*>0$,
\begin{equation}\label{eq:Vw}
\begin{aligned}
V^w&=(1-\theta)V_{z^\ast \alpha^\ast, z^\ast}(0)=\frac{K(1-\theta)(1+\beta)}{2} (z^\ast)^{-\eps},
\end{aligned}
\end{equation}
which is positive as granted.

Up to this point, we have not used the explicit form of $\kappa(z)$. It turns out that with our choice of $\kappa(z)=kz^\delta$, \eqref{eq:z*0}-\eqref{eq:Vw} can be significantly simplified, leading to a complete characterization of the equilibrium.

\begin{theorem}\label{thm:manager}
Let parameters $0\le \eps\le 1$, $\beta\in [0,1)$, $\kappa_0\ge 0$ as well as $K, p, \theta, k, \delta, c>0$ be given.
\begin{itemize}
\item[(i)] If $K(1+p)\theta\le \kappa_0$, the unique equilibrium for the two-layer mean field game is $(z^\ast, \alpha^\ast)=(0,0)$.
\item[(ii)] If $\left[1-\frac{\kappa_0}{K\theta}\right]\delta \ge \frac{(2-\eps)p}{1+p}$ which implies $K(1+p)\theta> \kappa_0$, the unique equilibrium team size is
\begin{equation}\label{eq:z*m}
z^\ast_m=\left[\frac{K\theta p (2-\eps)}{k\delta (1+p)}\right]^{\frac{1}{\delta}}>0.
\end{equation}
The associated equilibrium effort, the value for the team manager and each worker are given respectively by
\[\alpha_m^\ast(r)=\frac{K(1+p)(1-\beta)(1-\theta)}{2c}(z^\ast_m)^{1-\eps}(1-r)^p,\]
\begin{equation}\label{e552} V^m=\frac{K\theta [p(\eps+\delta-2)+\delta]}{\delta(1+p)}-\kappa_0 \ge 0,\end{equation}
\[V^w=\frac{K(1-\theta)(1+\beta)}{2}(z^\ast_m)^{-\eps}.\]
\item[(iii)] In all other cases, an equilibrium does not exist.
\end{itemize}
\end{theorem}

\begin{remark}
(a) Observe that in the case where a positive equilibrium team size exists, the equilibrium team size $z^\ast_m$, the associated manager's value $V^m$ and the representative worker's value $V^w$ are all independent of the cost parameter $c$. Similar phenomenon can be observed in the central planner's case (Sec~\ref{sec:cp}) and the partnership case (Sec~\ref{sec:partnership}) as well. This is a feature of the Poisson model and relative performance criteria which is in line with \cite{Nutz-Zhang.19}. A larger $c$ implies smaller effort. But when every co-worker and competing teams do so in equilibrium, the relative contribution $\alpha/\bar\alpha$ and the team's rank $\rho(\tau)$ remain unchanged, and the reduced cost per unit time for each worker is exactly offset by the the increased completion time $\tau$.

(b) It is also interesting to note that the equilibrium team size $z^\ast_m$ and the manager's value $V^m$ are independent of the base salary weight $\beta$, while the equilibrium effort $\alpha^*$ of each representative worker is decreasing in $\beta$. If we take $\beta$ as a control so that the team manager is responsible for designing the reward contract to distribute the weight between the base salary and the bonus income for each worker, it follows that a higher bonus income (smaller $\beta$) induces higher effort. However, this is at the expense of a smaller worker's value $V^w$. In addition, since $V^m$ is independent of $\beta$, the team manager has no incentive to raise the proportion of the bonus income. This is also related to the fact that we have a purely rank-based competition between teams where larger effort does not leads to larger expected reward if all other teams raise their intensity in equilibrium.

(c) Finally, we see from \eqref{eq:z*m} that in case (ii), the equilibrium team size $z^*_m$ is increasing in $\theta$, the team manger's share of the reward. As one would expect, $V^m$ given in \eqref{e552} is increasing in $\theta$, which follows from
$$\delta\geq\left[1-\frac{\kappa_0}{K\theta}\right]\delta \ge \frac{(2-\eps)p}{1+p}\implies (\eps+\delta-2)p+\delta\geq 0.$$
The worker's value $V^w$ is decreasing in $\theta$, but only for $\theta\ge \frac{\kappa_0\delta (1+p)}{K[\delta +(\delta+\eps-2)p]}$ which ensures the existence of a positive equilibrium team size. When $\theta$ is very small, we fall into case (i) where the team manager has no incentive to assemble the team, giving each worker zero value.
\end{remark}

\subsection{Optimal Team Size by A Central Planner}\label{sec:cp}
In this section, we consider another interesting problem when the size of all (homogeneous) teams is decided by a central planner of the whole society instead of each team manager. To this end, we look for an optimal team size rather than an equilibrium team size.
The assumption that the central planner takes the role to determine the unified team size can be seen as an example of the planned economy, in which each team represents a state-owned company. It is sometimes important for the government to wisely control the size of these large firms to benefit each individual's welfare in the society. This section provides a tractable model in the mean field sense for such a centralized optimization problem.

In the central planner's problem, the crux of the matter is to look for a unified team size so that each individual worker's average welfare in the team can be maximized as a static optimization problem:\footnote{One could also consider other criteria for the central planner, such as minimizing a given quantile of the completion time distribution or maximizing the total welfare of a team.}

\[V^c:=\sup_{z\ge 0} \left\{1_{\{z>0\}} \left(V_{z\alpha_z,z}(0)-\frac{\kappa_0+\kappa(z)}{z}\right)\right\},\]
where $V_{z\alpha_z,z}(0)$ is given in \eqref{EQM-V0}.

Let us denote by $h(z)$ the function to be maximized above. By \eqref{EQM-V0}, we get
\[h(z)
=1_{\{z>0\}}\left(\frac{K(1+\beta)}{2} z^{-\eps}-\kappa_0 z^{-1}-k\,z^{\delta-1}\right).\]

The next result on the optimal team size follows by straightforward computation and its proof is hence omitted.

\begin{theorem}\label{thm:socialplan}
Let parameters $\kappa_0\ge 0$, $\beta\in [0,1)$ and $K, p, k, \delta, c>0$ be given.

\textit{Case I}: $\eps=0$ (public good allocation scheme). In this case,
we have the following:
\begin{itemize}
\item If $0<\delta<1$, there is no optimal team size.
\item If $\delta=1$ and $\kappa_0>0$, then $z^*=0$ is the unique optimal team size if $\frac{K(1+\beta)}{2}\le k$ and there is no optimal team size if $\frac{K(1+\beta)}{2}> k$.
\item If $\delta=1$ and $\kappa_0=0$, we have the subcases:
\begin{itemize}
\item If $\frac{K(1+\beta)}{2} <k$, $z^*=0$ is the unique optimal team size;
\item If $\frac{K(1+\beta)}{2} =k$, any $z^*\geq 0$ is an optimal team size;
\item If $\frac{K(1+\beta)}{2}>k$, any positive team size is an optimal team size.
\end{itemize}
\item If $\delta>1$ and $\kappa_0=0$, there is no optimal team size.
\item If $\delta>1$ and $\kappa_0>0$,
let us define
\begin{align}
&z_c^\ast:=\left[\frac{\kappa_0}{k(\delta-1)}\right]^{1/\delta}\label{central-1}\\
\text{and}\ \ &V^c:=\frac{K(1+\beta)}{2} -\frac{\kappa_0\delta}{\delta-1} \left[\frac{\kappa_0}{k(\delta-1)}\right]^{-1/\delta}.\label{central-val-1}
\end{align}
Then we have:
\begin{itemize}
\item If $\frac{K(1+\beta)}{2}>\frac{\kappa_0\delta}{\delta-1} \left[\frac{\kappa_0}{k(\delta-1)}\right]^{-1/\delta}$, $z^\ast_c$ defined in \eqref{central-1} is the unique positive optimal team size and the intra-team equilibrium control is given by $\alpha_c^*(r)=\frac{K(1+p)(1-\beta)}{2c}z_c^* (1-r)^p$. The function $V^c>0$ defined in \eqref{central-val-1} is the associated value of a representative worker;
\item If $\frac{K(1+\beta)}{2}=\frac{\kappa_0\delta}{\delta-1} \left[\frac{\kappa_0}{k(\delta-1)}\right]^{-1/\delta}$, $z^*=0$ is another optimal team size in addition to $z^\ast_c$ given in \eqref{central-1};
\item If $\frac{K(1+\beta)}{2}<\frac{\kappa_0\delta}{\delta-1} \left[\frac{\kappa_0}{k(\delta-1)}\right]^{-1/\delta}$, $z^*=0$ is the unique optimal team size.
\end{itemize}
\end{itemize}

\textit{Case II}: $\eps=1$ (budget allocation scheme). In this case, we have the following:
\begin{itemize}
\item If $\kappa_0< \frac{K(1+\beta)}{2}$, then $h(0+)=\infty$ and there is no optimal team size.
\item If $\kappa_0\ge \frac{K(1+\beta)}{2}$, then $z^*=0$ is the unique optimal team size.
\end{itemize}
\end{theorem}

\begin{remark}\label{central-re-1}
In the budget allocation scheme ($\eps=1$), we see that the central planner prefers not to assemble the teams as either there is no optimal team size or the unique optimal solution is zero size. This can be explained by the fact that the reward to each worker under the budget allocation scheme is dominated by the effort cost and therefore there is no sufficient incentive for the worker to join the team in this case.

In Theorem \ref{thm:socialplan}, we have several parameter regimes under which there is no optimal team size because the target function $h(z)$ attains its maximum (possibly $\infty$) as $z\rightarrow 0$ (for instance, when $\kappa_0<\frac{K(1+\beta)}{2}$ under the budget allocation scheme $\eps=1$) or as $z\rightarrow \infty$ (for instance, when $0<\delta<1$ or
when $\delta=1$, $\kappa_0>0$ and $\frac{K(1+\beta)}{2}>k$ under the public good allocation scheme $\eps=0$). In reality, however, we usually have some lower and upper bounds for the value of $z$ in view that $z$ stands for the number of units of certain large population base. To ensure that the mean field interactions can be applied, the population within each team needs to be sufficiently large, which requires that the value of $z$ cannot be too small when the population base is chosen. On the other hand, when the population within a team is very large, it is natural to increase the population base instead of the units so that the value of $z$ usually lies in some reasonable finite range. Therefore, it is by no means restrictive in applications to exclude these parameter regions when there is no optimal team size. In our numerical examples in Section 4, we will only focus on the parameter values when an optimal team size exists.
\end{remark}

\begin{remark}
Similar to two main allocation schemes in Theorem \ref{thm:socialplan}, one can consider the mixed allocation scheme that $0<\eps<1$. Direct computations can show that there are only two cases with a unique non-trivial optimal team size:
\begin{itemize}
\item[(a)] $0<\eps,\delta<1$, $\eps+\delta< 1$ and $\kappa_0=0$. In this case, the unique optimal team size and the associated value of a representative worker are given by
\begin{align*}
&z_c^*:=\left(\frac{\eps K(1+\beta)}{2(1-\delta)k} \right)^{\frac{1}{\delta+\eps-1}},\\
\text{and}\ \ &V^c:= \frac{K(1+\beta)}{2} \left(\frac{\eps K(1+\beta)}{2(1-\delta)k} \right)^{\frac{-\eps}{\delta+\eps-1}}-k\left(\frac{\eps K(1+\beta)}{2(1-\delta)k} \right)^{\frac{\delta-1}{\delta+\eps-1}}.
\end{align*}

\item[(b)] $0<\eps<1$, $\delta\geq 1$, $\kappa_0>0$ and $\frac{K(1+\beta)}{2}>k+\kappa_0$.  In this case, the unique optimal team size $z_c^*$ is given by the unique positive root of the algebraic equation
\begin{align*}
\frac{-\eps K(1+\beta)}{2}z^{1-\eps}+\kappa_0-k(\delta-1)z^{\delta}=0,
\end{align*}
and the value of a representative worker is given by
$$V^c:=\kappa_0\left(\frac{1}{\eps}-1\right)\frac{1}{z_c^*}-k\left(1+\frac{\delta-1}{\eps}\right)(z_c^*)^{\delta-1}>0.$$
\end{itemize}

Recall in Theorem \ref{thm:socialplan} that the optimal team size $z^*_c$ is constant in $\beta$. This is because when $\eps=0$, the equilibrium reward\footnote{Here we refer to $V_{z\alpha_z, z}(0)$ as ``equilibrium reward" to separate it from size-related costs in the definition of $V^c$. It should be understood that $V_{z\alpha_z, z}(0)$ includes the cost of effort.} $V_{z\alpha_z, z}(0)$ each worker receives is independent of the team size $z$ (see \eqref{EQM-V0}). So the optimal team size is solely determined by the cost of building the team, while $\beta$ affects the reward but not the cost. In contrast, in the above cases (a) and (b) where $\eps>0$, the quantity $V_{z\alpha_z, z}(0)$ is decreasing in $z$. Because $V_{z\alpha_z, z}(0)$ depends positively on $\beta$, as $\beta$ increases, the central planner is motivated to choose a smaller team size so that each worker can get a larger share of the increased reward.

It is worth pointing out that in our model, the inter-team competition has a fixed reward pie that is shared purely based on relative performance. If absolute performance or effort level also enter into the reward scheme, then the dependence of $V_{z\alpha_z, z}(0)$ on $\beta$ will likely be more complicated, since a larger $\beta$ leads to more serious free-rider problem, which apart from reducing the cost of effort, will also have a negative impact on the team's reward.
\end{remark}

\subsection{Equilibrium Team Size with Partnership}\label{sec:partnership}

In this section, we analyze the model when all teams are assembled neither by team managers nor a central planner, but directly by a group of workers joining together as a partnership. The team size is determined by public voting to maximize each worker's benefit. Each team still competes with all other teams though the rank-based reward, and the goal is to find the equilibrium team size in the framework of partnership for each team.

We focus on the static decision making over the team size using the representative worker's expected payoff instead of the payoff by the team manager. We again focus on a representative team $i$ and assume that all other teams choose the same jump intensity $\lambda\in\mathcal{A}$ or $\lambda\equiv 0$. After public voting in the partnership setting, a representative team member from team $i$ needs to solve an optimization problem
\begin{align*}
\sup_{z\ge 0}\left[1_{\{z>0\}}\left(V_{\lambda,z}(0)-\frac{\kappa_0+\kappa(z)}{z}\right) \right],
\end{align*}
where $V_{\lambda,z}(0)$, given by \eqref{DPE2soln} for $\lambda\in\mathcal{A}$ or \eqref{eqq2} for $\lambda\equiv 0$, represents the intra-team value of a representative worker in team $i$.

Let us denote the objective function
\begin{align*}
J^p(z;\lambda):=1_{\{z>0\}}\left(V_{\lambda,z}(0)-\frac{\kappa_0+\kappa(z)}{z}\right).
\end{align*}

We next give the definition of a Nash equilibrium in the setting of two-layer mean field game with partnership.
\begin{definition}
A pair $(z^\ast,\alpha^\ast)\in \R_+\times \cA$ is said to be an equilibrium for the two-layer mean field game with partnership if
\[ z^* \in \argmax_{z\ge 0} J^p(z, z^\ast \alpha^\ast)\quad \text{and}\quad \alpha^\ast=\alpha_{z^\ast}.\]
That is, $z^\ast$ is the optimal size determined by public voting given that all other teams choose the size $z^*$ and the individual control $\alpha^\ast$; and $\alpha^\ast$ is the intra-team equilibrium control given that all teams including the representative one have size $z^\ast$. We refer to such a
$z^\ast$ as an equilibrium team size with partnership and such an $\alpha^\ast$ as the associated equilibrium effort.
\end{definition}

By its definition, $z^\ast$ is an equilibrium team size if and only if either
\begin{align*}
z^*=0\ \ \ \text{and}\ \ \ \sup_{z>0}\left(\frac{1+\beta}{2} G_z(0)-\frac{\kappa_0+kz^{\delta}}{z} \right)\leq 0,
\end{align*}
or
\begin{align*}
z^*>0\ \ \ \text{and}\ \ \ z^\ast  \in \argmax_{z\geq 0}\ \left\{1_{\{z>0\}}\left(V_{\lambda^*,z}(0)-\frac{\kappa_0+kz^{\delta}}{z}\right)\right\},
\end{align*}
where
\begin{align*}
\lambda^\ast(x)=z^\ast \alpha_{z^\ast} =\frac{(1-\beta)K(1+p)}{2c}(z^\ast)^{2-\epsilon}(1-x)^p.
\end{align*}

In the first case, let us recall that $G_z(0)=K(1+p)\cdot z^{-\epsilon}$. In the second case with $z^*>0$, in view of \eqref{DPE2soln}, to find the equilibrium team size $z^*$ is equivalent to maximize the function
\begin{align}\label{partnerH}
H(z;z^*):=&J^p(z;\lambda^*)\notag\\
=& -\frac{\kappa_0+k z^{\delta}}{z}+\frac{1-\beta^2}{4c}K^2 (1+p)^2z^{2-2\eps}\notag\\
&\int_0^1\frac{(1-x)^{2p-1}}{\lambda^*(x)}\exp\left(-\frac{K(1+p)(1-\beta)}{2c} z^{2-\eps}\int_0^x\frac{(1-y)^{p-1}}{\lambda^*(y)}dy\right)\,dx\notag\\
=&\frac{K(1+p)(1+\beta)z^{2-2\varepsilon}}{2(z^*)^{2-\varepsilon}} \frac{1}{p+(\frac z{z^*})^{2-\varepsilon}}-\frac{\kappa_0+kz^\delta}{z}.
\end{align}

As the target functions are quite complicated comparing with the model with team managers or a central planner, we only provide some sufficient conditions on model parameters for the existence and uniqueness of the equilibrium in the model with partnership. 
We have the following results on the equilibrium pair $(z^\ast, \alpha^\ast)$.

\begin{theorem}\label{thmpartner}
Let parameters $\kappa_0\ge 0$, $\beta\in [0,1)$ and $K, p, k, \delta, c>0$ be given. \\
\ \\
\textit{Case I}: $\varepsilon=0$ (public good allocation scheme). We have the following results:
\begin{itemize}
\item[(i)] Assume $\kappa_0=0$. Then $z^*=0$ is an equilibrium team size if and only if $\delta=1$ and $\frac{(1+\beta)}{2} K(1+p)\leq k$.
\item[(ii)] Assume $\kappa_0>0$ and $\delta\in(0,1]\cup[2,\infty)$. Then $z^*=0$ is an equilibrium team size if and only if $$\frac{2^\delta(\kappa_0)^{\delta-1}k\delta^{\delta}}{(1+\beta)^{\delta}K^{\delta}(1+p)^{\delta}(\delta-1)^{\delta-1}}\geq 1.$$
\item[(iii)] Assume $\delta\geq 3$ and $p\geq 1/3$. Let $z_p^*$ be the unique positive solution to the algebraic equation
\begin{align*}
\frac{pK(1+\beta)}{1+p}z+k(1-\delta)z^{\delta}+\kappa_0=0.
\end{align*}
Let us define the function
\begin{align*}
V^p:=\frac{K(1+\beta)}{2}-\kappa_0(z^\ast_p)^{-1}-k(z^\ast_p)^{\delta-1}.
\end{align*}
Then $z_p^*$ is the unique positive equilibrium team size if and only if $V^p\geq 0$; in this case, the corresponding equilibrium individual control is $\alpha_p^*(r)=\frac{K(1+p)(1-\beta)}{2c}z^\ast_p(1-r)^p$, and $V^p$ is the associated value of each representative worker. 

\end{itemize}
\noindent
\textit{Case II}: $\varepsilon=1$ (budget allocation scheme). We have the following results: 
\begin{itemize}
\item[(i)] $z^*=0$ is an equilibrium team size if and only if $\frac{(1+\beta)}{2} K(1+p)\leq \kappa_0$.
\item[(ii)] Assume $\delta\geq 2$, $2(1+\delta)>(1+p)^2$ and $\frac{K(1+\beta)}{2(1+p)}<\kappa_0$. Define the positive constant $z_p^*$ by
\begin{align}\label{equisol-size}
z_p^*=\left(\frac{\kappa_0-\frac{K(1+\beta)}{2(1+p)}}{k(\delta-1)}\right)^{\frac{1}{\delta}}.
\end{align}
and the function
\begin{align*}
V^p:=\frac{K(1+\beta)}{2}(z^\ast_p)^{-1}-\kappa_0(z^\ast_p)^{-1}-k(z^\ast_p)^{\delta-1}.
\end{align*}
Then $z_p^*$ is the unique positive equilibrium team size if and only if $V^p\geq 0$, which is also equivalent to
\begin{equation}\label{eqq1}
\kappa_0\leq \frac{K(1+\beta)}{2(1+p)}\cdot \frac{1+(\delta-1)(1+p)}{\delta};
\end{equation}
in this case, the equilibrium individual control is given by $\alpha_p^*(r)=\frac{K(1+p)(1-\beta)}{2c}z_p^*(1-r)^p$, and  $V^p$ is the associated value of each representative worker. 
\end{itemize}
\end{theorem}

\begin{remark}
Note the ranges of $\kappa_0$ in \eqref{eqq1} and Case II of Theorem \ref{thm:socialplan}. It can be shown that
$$\frac{K(1+\beta)}{2(1+p)}\cdot \frac{1+(\delta-1)(1+p)}{\delta}<\frac{K(1+\beta)}{2}.$$
Therefore, in the case given by \eqref{eqq1} there is no optimal team size for the corresponding central planner problem. On the other hand, when $\eps=1$ and $\kappa_0\geq\frac{K(1+\beta)}{2}$ there is no positive equilibrium team size for the partnership problem by Case II of Theorem \ref{thm:socialplan}.
\end{remark}

It can be easily shown that in Case I (iii) of Thereom \ref{thmpartner}, both $z_p^*$ and the associated $V^p$ is increasing with respect to (w.r.t.) $\beta$; in Case II (ii) $z_p^*$ is decreasing and the associated $V^p$ is increasing w.r.t.\ $\beta$. Below we provide more general results regarding the monotonicity of $z_p^*$ and $V^p$ w.r.t. $\beta$.

\begin{proposition}\label{p1}
Let parameters $\delta>1, \eps\in[0,1]$, and $K, \kappa_0, k>0$ be given. Let $A$ be a connected interval in $[0,1]$. Suppose for any $\beta\in A$ there exists a positive equilibrium team size $z_p^*=z_p^*(\beta)$. Then $z_p^*$ is unique and monotone w.r.t. $\beta$ on $A$. Specifically, we have that
\begin{itemize}
\item if $2p-2p\eps-\eps>0$, then $z_p^*$ is increasing w.r.t. $\beta$;
\item if $2p-2p\eps-\eps=0$, then $z_p^*$ is independent of $\beta$;
\item if $2p-2p\eps-\eps<0$, then $z_p^*$ is decreasing w.r.t. $\beta$.
\end{itemize}
\end{proposition}

\begin{remark}
As $\eps$ gets larger, there is more division effect. Note that the function in \eqref{partnerH} is increasing w.r.t.\ $\beta$. As a result, for relatively large $\eps$, there is more incentive to choose a small size in order to get a larger pie when $\beta$ is large. That is, $z_p^*$ is more likely to be decreasing w.r.t.\ $\beta$ when $\eps$ is large, which is the implication of the above result.
\end{remark}

\begin{proposition}\label{p2}
Let parameters $\delta>1, \eps\in[0,1]$, and $K, \kappa_0, k>0$ be given. Suppose  $z_p^*=z_p^*(\beta)$ is the unique positive equilibrium team size for any $\beta\in A\subset[0,1]$, where $A$ is a connected interval. From \eqref{partnerH} we define the value for each team member under $z_p^*$ by
$$V^p(\beta,z_p^*(\beta)):=V^p(\beta,z_p^*):=\frac{K(1+\beta)}{2}(z_p^*)^{-\eps}-\frac{\kappa_0+k(z_p^*)^\delta}{z_p^*}.$$
If $2p-2p\eps-\eps\leq 0$, or $2p-2p\eps-\eps>0$ and $\frac{p(2-\eps)}{(p+1)(\delta+\eps-1)}\leq 1$, then $\beta\mapsto V^p(\beta,z_p^*(\beta))$ is increasing on $A$.
\end{proposition}

\begin{remark}
As $\beta\in A$ gets larger, there is less competition in the same team and thus less effort cost and more reward.
The only possibility of $V^p$ getting smaller would be due to the change of team size $z_p^*(\beta)$. The above result indicates that the change of $z_p^*$ (as $\beta\in A$ gets larger) would either lead to an increase in $V^p$, or be dominated by the increase of $V^p$ due to less competition.
\end{remark}

\section{Numerical Examples}\label{sec-4}

In this section, we choose a representative worker's value under the equilibrium or optimal team size
as the metric to compare which model works better for the team size decision making problem. Our numerical figures illustrate that all models have their pros and cons depending on the model parameters.

Recall that $V^w,V^c,V^p$ denote a worker or a regular team member's value in Sections 3.1-3.3 under the equilibrium or optimal team size respectively. From their definitions, we always have the order that
$$V^c\geq V^p.$$
However, the comparison between $V^w$ and $V^c$ or $V^w$ and $V^p$ is not straightforward and can lead to several interesting economic insights. In particular, we will fix parameters $K, p, \eps,\delta, k, \kappa_0$ and treat either $\theta$ or $\beta$ as the variable to illustrate its quantitative impact on each worker's value and the equilibrium or optimal team size. We can also numerically split the regions of $\theta$ or $\beta$ such that the value functions may dominate each other in different regions to make the comparison analysis among three different models.

\subsection{Example 1}

We first choose $\theta\in (0,1)$, i.e. the proportion of team's reward allocated to the team manager, as the changing variable and set other model parameters $K=20/3$, $\delta=4, p=2, k=1,\beta=0.4$ and $\kappa_0=2$. We aim to plot comparison figures under the public good allocation scheme $\eps=0$.

In the model with team managers, it follows from Theorem \ref{thm:manager} $(i)$ that $K(1+p)\theta\leq \kappa_0$ becomes $\theta\leq 0.1$. Therefore, on the variable region $\theta\in (0, 0.1]$, the equilibrium team size $z^*_m=0$. That is, the team manager will not assemble the team and the value function of the worker is $V^w=0$. On the other hand, by Theorem \ref{thm:manager} $(ii)$, we have that $\left[1-\frac{\kappa_0}{K\theta}\right]\delta \ge \frac{(2-\eps)p}{1+p}$ is satisfied if and only if $\theta \ge 0.45$. That is, for the variable region $\theta\in [0.45,1)$, the equilibrium team size is given by $z^\ast_m=\left(\frac{20}{9}\theta\right)^{\frac{1}{4}}$ and the value function of the worker's is $V^w=\frac{14}{3}(1-\theta)$. For the remaining region $\theta\in(0.1, 0.45)$, the equilibrium team size does not exist according to Theorem \ref{thm:manager} $(iii)$.

In the model with a central planner, the condition $\frac{K(1+\beta)}{2}>\frac{\kappa_0\delta}{\delta-1} \left[\frac{\kappa_0}{k(\delta-1)}\right]^{-1/\delta}$
is satisfied by our chosen parameters in the case $\eps=0$ in Theorem \ref{thm:socialplan}, which gives the optimal team size by $z_c^\ast\approx 0.904$ and the corresponding optimal value of a representative worker is computed by $V^c\approx 1.716$.

In the third model with partnership within each team, as $\delta=4$ and $p=2$, by item $(iii)$ in the case $\eps=0$ of Theorem \ref{thmpartner}, the equilibrium team size $z_p^*$ is the unique positive root of the algebraic equation
\begin{align*}
-3z^4+\frac{56}{9}z+2=0
\end{align*}
which gives that $z_p^*\approx 1.368$ and the value of the worker is $V^p\approx 0.644$.

Based on the computations above, we plot and compare in Figures 1 and 2 the three values
$V^w$, $V^c$ and $V^p$ and the three different team sizes $z_m^*$, $z_c^*$ and $z_p^*$ as functions of the variable $\theta\in (0,1)$, under the public good allocation scheme $\eps=0$.

$$
\begin{array}{cc}
\includegraphics[height=1.7in]{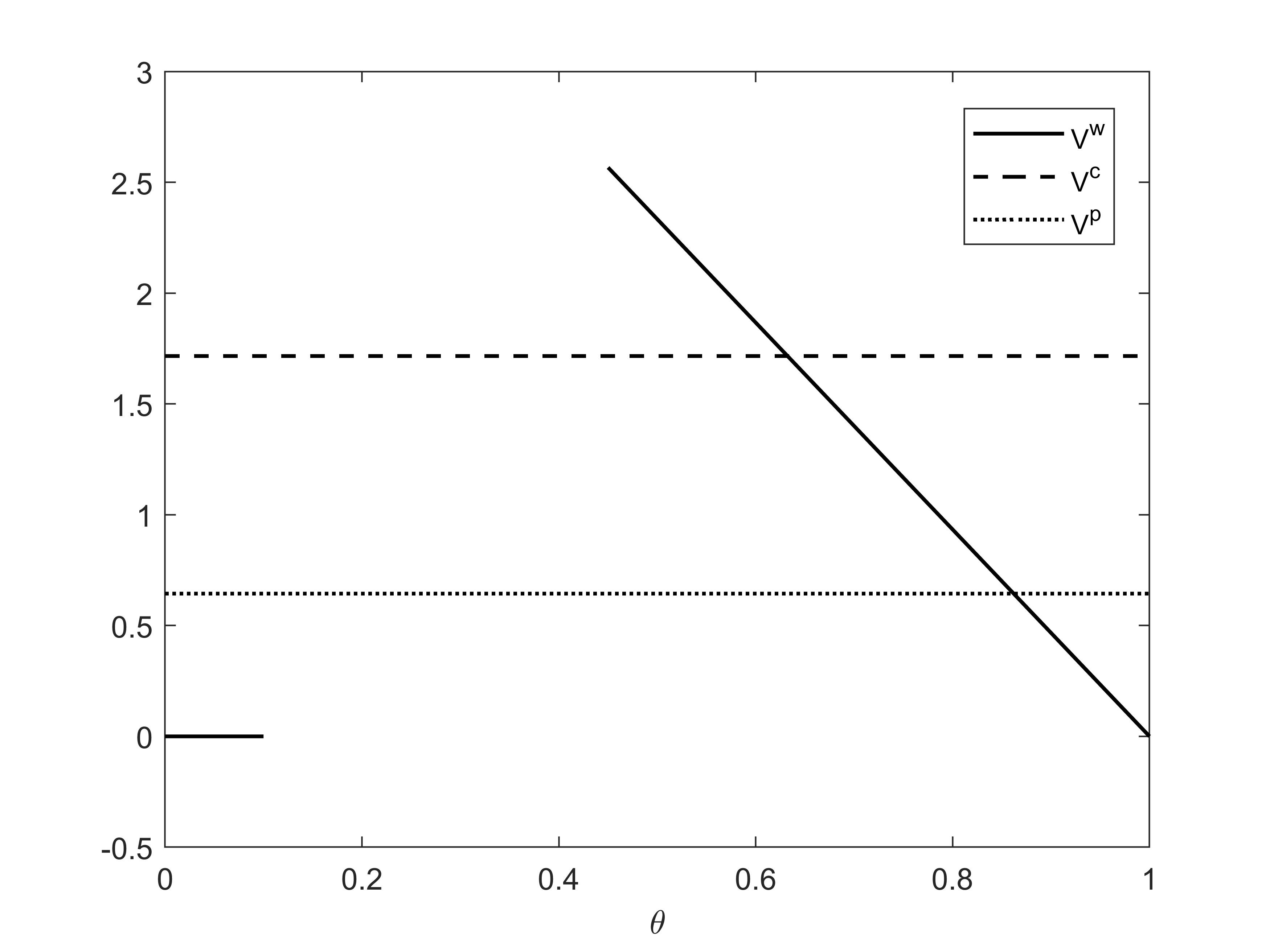}\quad\quad\includegraphics[height=1.7in]{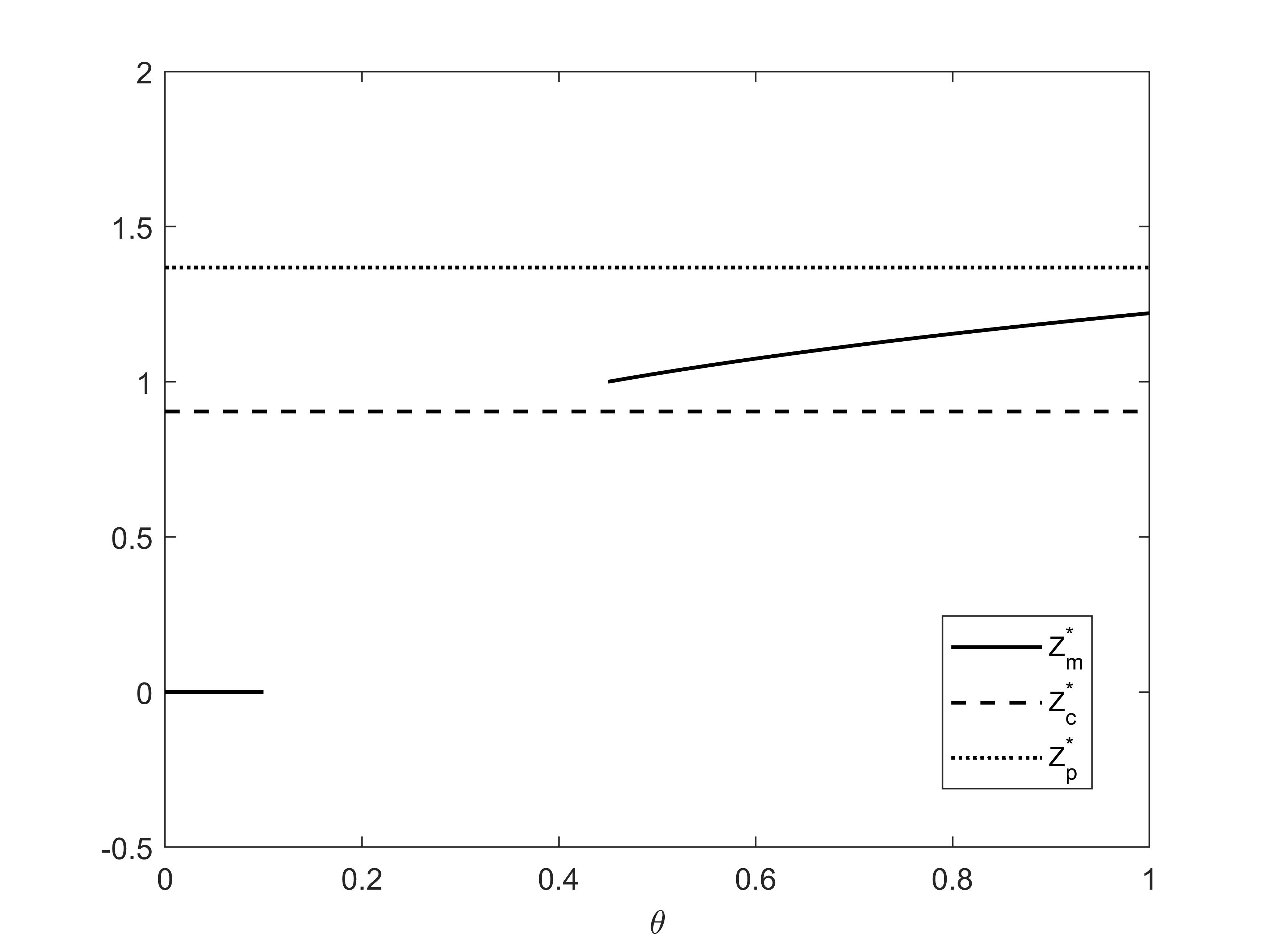}\\
\mbox{\small{Figure 1}}\hspace{2in}\mbox{\small{Figure 2}}
\end{array}
$$


From Figure 1, we can see that the team manager has no incentives to assemble the team (either because the equilibrium team size is zero or there is no equilibrium team size) to work on his behalf if he can only earn a small proportion $\theta< 0.45$ from the team's total gain $K(1+p)(1-\rho(\tau))^p$ for the given parameters. When the team manager is allowed to acquire a higher proportion $\theta\geq 0.45$ from the team's total gain, both the team manager and the representative worker become motivated to take part in the teamwork with different roles. Within a reasonable region $\theta\in [0.45, 0.632]$, it becomes a win-win situation between the team manager and the hired worker as both of them can attain high values $V^m$ and $V^w$. More importantly, by comparing three curves in Figure 1, we can see that $V^w\geq V^c$ on $\theta\in [0.45, 0.632]$, which implies that the cost sharing mechanism in the design of the first model totally beats the wage paid to the team manager and it is therefore more beneficial on the top layer to hire a team manager to run the team business than other two models. Similarly, for $\theta\in [0.45, 0.862]$, the value function $V^w$ in the model with team managers outperforms the value function $V^p$ in the model with voluntary partnership.

However, if the team manager is getting more greedy and aims to eat a larger chunk of the profit pie, the remaining gain to share among team workers becomes more limited. This is illustrated by the curve $V^w$ for $\theta\geq 0.45$ that $V^w$ is decreasing in $\theta$. Moreover, as observed from Figure 1, as $\theta$ surpasses the threshold $0.632$, the dominance relationship between $V^w$ and $V^c$ is overturned and each worker is better off if the team size is assigned by a central planner so that no extra wage is paid from the team's account. In the language of economics, we can interpret the first model with team managers as an example of market economy, in which the size of each firm is decided by the manager based on market competition. On the other hand, we can also view the second model with a central planner as a toy model of the planned economy, in which the government has the right to operate the firm size of large state-owned enterprises. Our numerical example here illustrates from the perspective of mean field competitions that under the public good allocation scheme, the market economy is more conducive to each worker's value when the senior management wage stays in a reasonable range. When the salary of the team manager is too high, the centralized management will become more preferable by the worker. Similarly, for $\theta>0.862$, we have that $V^w$ falls below $V^p$, which indicates that the role of a team manager becomes unnecessary as the partnership organization is more beneficial to each worker and the wage amount allocated to the team manager dominates the effect of team size costs.

It is also interesting to see from Figure 2 that the larger the equilibrium team size is, the smaller the value $V^w$ becomes, despite that here we have no division effect. This shows the possible adverse effect by the team size to each worker's profit. We can observe that as the team manager gets a higher salary (as $\theta$ increases), he is more motivated to enlarge the team size by sacrificing the worker's value. Comparing with $z_c^*$ and $z_p^*$, we can also see that the role of a central planner will not only enhance each worker's profit, but it will also yield leaner settings and downsize the work force by the centralized management.

\subsection{Example 2}

In this example, we choose $\beta\in [0,0.8)$, i.e., the proportion that the rank-based reward is split between the fixed salary and the performance-based bonus for each worker, as the changing variable and set other model parameters the same as in Example 1 that $K=\frac{20}{3}$, $\delta=4, p=2, k=1,\theta=0.5$ and $\kappa_0=2$. We plot the comparison figures for the budget allocation scheme $\eps=1$.

In the first model with team managers, the condition $\left[1-\frac{\kappa_0}{K\theta}\right]\delta \ge \frac{(2-\eps)p}{1+p}$ in Theorem \ref{thm:manager} $(ii)$ is always satisfied by our chosen parameters. That is, for any $\beta\in [0,0.8)$, the equilibrium team size determined by the team manager is given by $z^\ast_m\approx 0.863$, and the value of a representative worker is given by the linear function $V^w=1.931(1+\beta)$.

In the second model with a central planner, for the case $\eps=1$ in Theorem \ref{thm:socialplan}, we have $\kappa_0<\frac{K(1+\beta)}{2}$ for any $\beta\in [0,0.8)$ and the optimal team size does not exist. We shall therefore skip its graph in the Figure.

At last, in the third model with partnership within each team, by item $(ii)$ in the case $\eps=1$ of Theorem \ref{thmpartner}, the conditions $\delta\geq 2$ and $2(1+\delta)>(p+1)^2$ are satisfied. The condition $\frac{K(1+\beta)}{2(p+1)}<\kappa_0$ is fulfilled if and only if $\beta\in [0,0.8)$. Therefore, for our chosen region $\beta<0.8$, its equilibrium team size is given explicitly by $z_p^*=\left(\frac{8}{27}-\frac{10}{27}\beta\right)^{\frac{1}{4}}$, and the value function of each worker is computed by $V^p=\left(\frac{4}{3}+\frac{10}{3}\beta \right)\left(\frac{8}{27}-\frac{10}{27}\beta\right)^{\frac{-1}{4}}-  \left(\frac{8}{27}-\frac{10}{27}\beta\right)^{\frac{3}{4}}$.

Under the budget allocation scheme $\eps=1$, we now compare the model with team manager and the model with partnership by plotting the curves of $V^w$ and $V^p$ for $\beta\in [0,0.8)$ in Figure 3. We also present two equilibrium team sizes $z_m^*$ and $z_p^*$ in Figure 4 for $\beta\in [0,0.8)$ as below.

$$
\begin{array}{cc}
\includegraphics[height=1.7in]{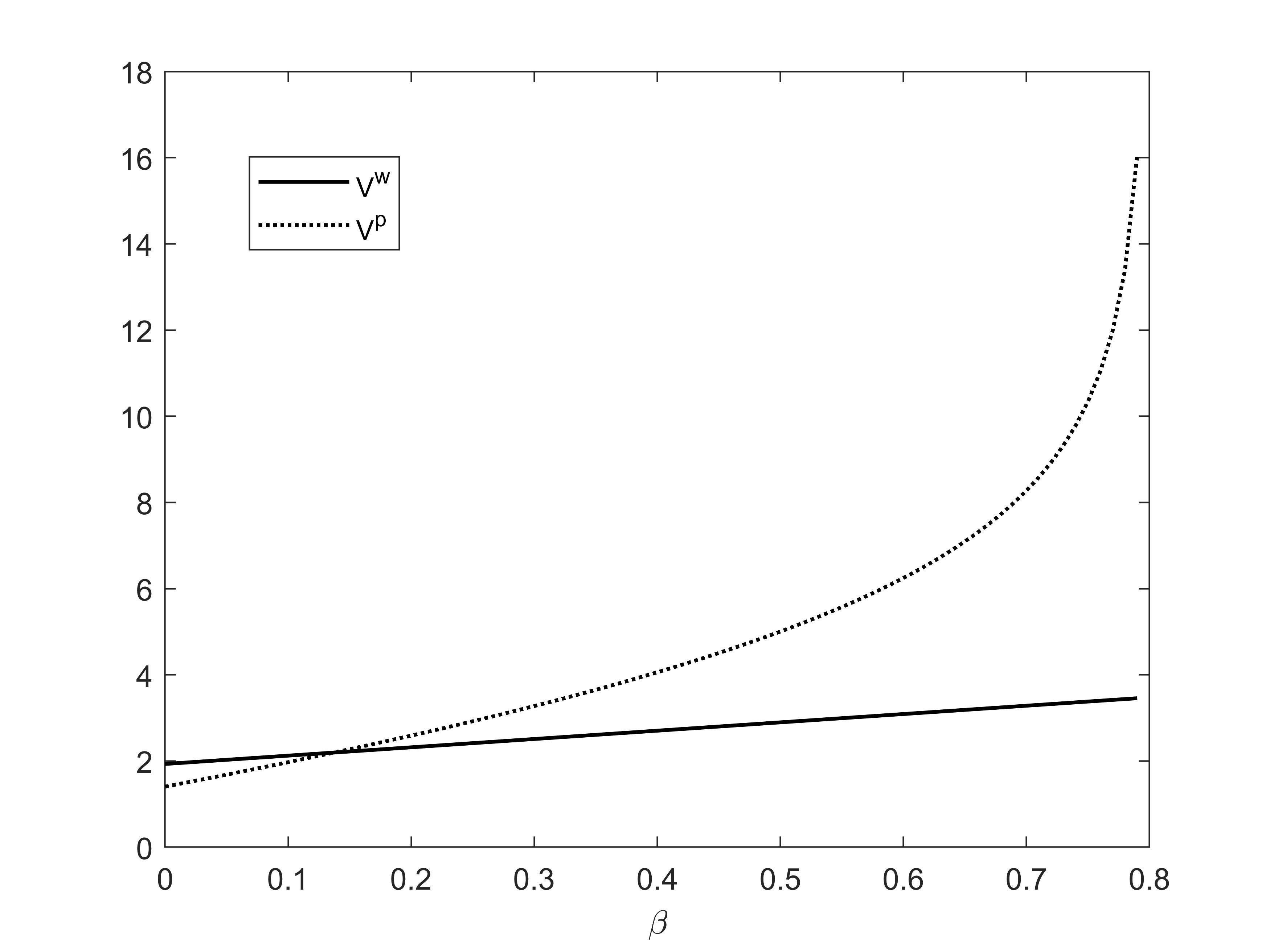}\quad\quad\includegraphics[height=1.7in]{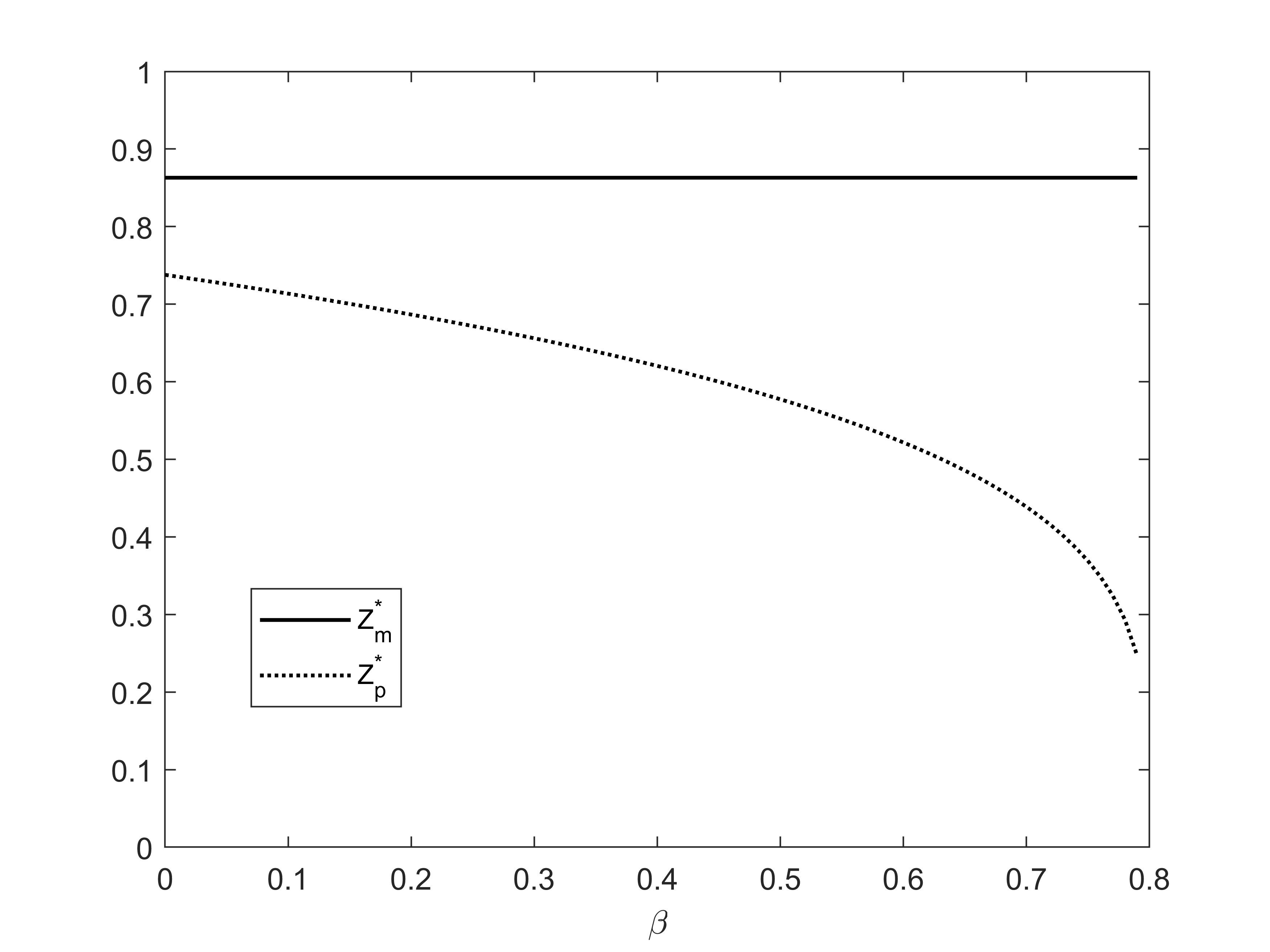}\\
\mbox{\small{Figure 3}}\hspace{2in}\mbox{\small{Figure 4}}
\end{array}
$$
%

First of all, we can see from Figure 3 and Figure 4 that $V^w$ and $V^p$ are increasing functions of $\beta$ and $z_p^*$ is a decreasing function of $\beta$, which are consistent with the remarks in Section 3.1 and Section 3.3. The economic insights on the impact by $\beta$ can be found therein.

For the region $\beta\in [0,0.1374]$ in Figure 3, we can see that $V^w>V^p$, which shows that in the budget allocation framework, the reward scheme using a larger performance-based bonus makes the model with team managers more competitive and appealing. To be precise, it is interesting to see that a larger bonus incentive (as $\beta$ is close to $0$) corresponds to a larger equilibrium team size  $z_p^*$ in the model of partnership. Figure $3$ illustrates that it is more wise for hard-working workers (who aim to receive a high bonus) in this scenario to look for and pay a team manager who can bear all the large team size costs for the team (as $z_p^*$ is large when $\beta$ is small) instead of their own partnership organization. As the value of $\beta$ increases, the fixed salary amount plays a more leading role in the reward scheme and the value $V^p$ grows more rapidly and eventually dominates $V^w$. This is consistent with some real life observations that if the rank-based reward towards the team is fixed and the performance-based bonus is very small to incentivize the worker, the role of a team manager becomes redundant. Workers will prefer to voluntarily gather to reduce the team size (as $z_p^*$ decreases in Figure 4) and tolerate the small size costs by themselves, but not to join a team in which the team manager may need to occupy a large proportion from the fixed reward pie.

On the other hand, as $\beta\rightarrow 0.8$ in Figure $3$, we can see that $V^p$ tends to a very satisfactory high value. However, from Figure 4, the price to pay to lift up $V^p$ is to reduce the team size $z_p^*$ to the extreme low value close to zero. Similar to Remark \ref{central-re-1}, to fit into our mean field formulation, it is not allowed to downsize $z_p^*$ too much as the base population is fixed at the beginning. That is, to take the advantage of the explicit intra-team equilibrium control and equilibrium team size $z_p^*$, we need to balance the payment to each worker between the fixed salary amount and the relative performance bonus by setting $\beta$ not too large under the budget allocation scheme. For example, by setting $\beta\approx 0.6$, we can guarantee that the reasonable team size $z_p^*\approx 0.522$ and the value function $V^p\approx 6.247$ is high enough to attract individual workers to gather as a self-organized team.

\section{Discussion on extension to heterogeneous cases}
In this section, we briefly discuss the extension of our results to heterogeneous cases. It is straightforward to generalize our results to heterogeneous population at the intra-team level. However, it becomes much more difficult at the inter-team level.

\subsection{Heterogeneity at intra-team level}

Let us consider the heterogeneity at the intra-team level; that is, within each team, members may have different cost of effort parameter $c$. We assume that the distribution of $c$ is supported on a compact subset of $\mathbb{R}_{++}$ and is the same across all teams regardless of the team sizes (e.g., $50\%$ of the members have $c=1$ and the rest $50\%$ has $c=2$, no matter what the team size $z$ is).

Recall that $(J, \mathcal{J}, \nu)$ is the finite non-negative measure space on which the the continuum of players of a representative team is defined. 
Consider member $j$ of this team with cost parameter $c_j$.
This member's control, which is given by \eqref{alphaz} in the homogeneous case, becomes:
\begin{equation}\notag
\alpha_j(r)=\frac{(1-\beta)zG_z(r)}{2c_j}.
\end{equation}
Hence, the average\footnote{Here and in the sequel, the average is taken in the space $(J, \mathcal{J}, \nu)$.} control of this team is given by
$$\bar\alpha(r):=\frac{1}{\nu(J)}\int \alpha_j(r) \nu(dj)=\frac{(1-\beta)zG_z(r)}{2\bar c},$$
where $\bar c$ is such that
$$\frac{1}{\bar c}=\frac{1}{\nu(J)}\int \frac{1}{c_j} \nu(dj) .$$
Moreover, by averaging the ODE, one can easily show that the average value function of team members is still given by \eqref{DPE2soln} with $c$ replaced by $\bar c$. Then in the manager's problem, we can just replace $\tilde c=c/(1-\theta)$ with $\tilde c=\bar c/(1-\theta)$ in the computation of equilibrium team sizes. (Note that Theorem \ref{thm:manager} does not change as it is independent of $\tilde c$.) The same applies to the central planner's problem in which the average team member's welfare is to be maximized. Finally, for the partnership problem, each member will prefer a different team size due to different cost coefficients. To reach a team decision, workers forming the partnership could select the team size that maximizes the average payoff. This again amounts to replacing $c$ by $\bar c$ in the computation and the result remains valid.

\subsection{Heterogeneity at inter-team level}


The situation becomes much more complicated when we consider heterogeneity at the inter-team level. To wit, consider the case with distinct $\beta$ and $c$ for different teams. Teams with smaller $\beta$ and $c$ work harder and tend to jump earlier. For each team, the probability of a jump occurring by time $t$ still satisfies \eqref{eq:rho}, where the jump intensity $\lambda$ is now team-specific.
Aggregating over the team space $(I, \mathcal{I}, \mu)$ and using the exact law of large numbers, we obtain that the proportion of teams that have jumped by time $t$ is given by
\[\rho(t)=\int_{i\in I} \rho_i(t) \mu(di)=\int_{i\in I} \int_0^t \lambda_i(\rho(s))(1-\rho_i(s))ds \mu(di).\]
Note that $\rho(t)$ can not be characterized by \eqref{eq:rho} with $\lambda$ replaced by an average intensity. 
As a result, dynamic programming using $\rho(t)$ as the state variable no longer works.

One possible remedy is to fix $\rho(t)$ as a (deterministic) input rather than a state variable in the single player's problem and consider time-dependent controls $\alpha(t)$ and value function
\[V_i(t)=V_i(t; \rho, z, \bar\alpha)=\sup_{\alpha}\E\left[G_z(\rho(\tau)) \cdot \left(\beta_i+(1-\beta_i)\frac{\alpha(\tau)}{\bar \alpha (\tau)}\right)-c_i \int_t^{\tau}\alpha^2(s)ds\right]\]
where $\tau=\inf\{s\ge t: \int_t^s z \bar\alpha(u)du=Z^i\}$. Assuming $\rho(\infty)=1$, then $V_i$ is expected to satisfy the dynamic programming equation
\[V_i'(t)+\sup_{\alpha} \left\{\left[G_z(\rho(t)) \cdot \left(\beta_i+(1-\beta_i)\frac{\alpha(t)}{\bar\alpha(t)}\right)-V_i(t)\right] \cdot z\bar\alpha(t) -c_i\alpha^2(t)\right\}=0, \quad V_i(\infty)=0.\]
Imposing the intra-team consistency condition
\[\bar\alpha(t)=\alpha^i_{\rho, z}(t)=\frac{(1-\beta_i)z G_{z}(\rho(t))}{2c_i},\]
we arrive at a first order ODE similar to \eqref{DPE2}:
\[V'_i(t)-\frac{1-\beta_i}{2c_i}z^2G_z(\rho(t)) V_i(t)+\frac{1-\beta_i^2}{4c_i}z^2G_z^2(\rho(t))=0,\quad V_i(\infty)=0.\]
As $\rho(t)$ is $[0,1]$-valued, the above equation can be easily solved, giving the equilibrium value function of members of team $i$ before project completion. The above can be made rigorous by working with a suitable class of admissible controls.

The difficulty lies in the inter-team equilibrium. Take the manager's problem as an example. As before, we replace the cost parameter $c_i$ by $\tilde c_i=c_i/(1-\theta)$ and, with a little abuse of notation, we write $\alpha^i_{\rho, z}(t)=(1-\beta_i)z G_{z}(\rho(t))/(2\tilde c_i)$. Also let
\[J_i^m(z; \rho)=1_{\{z>0\}}\left(\theta \mathbb{E}\left[ z^\eps G_z(\rho(\tau^i_{\rho,z}))\right]-\kappa_0-\kappa(z)\right),\]
where $\tau^i_{\rho, z}$ is the jump time of team $i$ when all of its members use control $\alpha^i_{\rho, z}$. An equilibrium for the two-layer mean field game in this setting becomes a pair $(\rho^*, \{z^*_i: i\in I\})$ such that
\[z_i^* \in \argmax_{z} J_i^m(z;\rho^*)
\]
and $\rho^*(t)=\int_{i\in I} \tilde \rho_i(t; \rho^*, z^*_i) \mu(di)$ where $\tilde \rho_i(t; \rho, z)$ is the solution to
\begin{equation*}
\tilde \rho_i(t)=\int_0^t z \alpha^i_{\rho, z}(t)(1-\tilde \rho_i(s))ds.
\end{equation*}
As opposed to the homogeneous case, an explicit characterization of $z^*_i$ is no longer available. Moreover, it is unclear whether the argmax is a singleton, which poses an open problem even for the abstract existence. 

To see this, let us assume for simplicity that each team manager can only choose team size $z$ in a compact subset $[z_{\min}, z_{\max}]$ of $\mathbb{R}_{++}$, and that there are only $M\in\N$ types of teams, with proportions $p_1, \ldots, p_M$. In what follows, with a little abuse of notation, we will index teams by their type $k\in\{1, \ldots, M\}$ instead of name $i\in I$.
Using that $\tau^k_{\rho, z}$ is the first jump time of an inhomogeneous Poisson process, we obtain that 
on $[z_{\min}, z_{\max}]$,
\begin{align*}
J^m_k(z;\rho)
&=\theta z^\eps \int_0^{\infty} \frac{1-\beta_k}{2\tilde c_k}z^2 G_z^2(\rho(s))\exp\left(-\int_0^s \frac{1-\beta_k}{2\tilde c_k}z^2 G_z(\rho(u))du \right)ds -\kappa_0-\kappa(z).
\end{align*}
Let us identify $\rho$ with the cdf of the elements in $\mathcal{P}([0,\infty])$, where we view $[0, \infty]$ as the one-point compactification of $[0, \infty]$ that is homeomorphic to $[0,1]$, and equip $\mathcal{P}([0,\infty])$ with the topology of weak convergence.
Treating $J^m_i$ as a mapping defined on $\mathcal{P}([0,\infty]) \times [z_{\min}, z_{\max}]$, it is continuous with respect to the product topology. By Berge's Maximum Theorem (see e.g. \cite{InfDimAnalysis}), the set-valued map
\[\Phi_k(\rho):= \argmax_{z\in [z_{\min}, z_{\max}]} J_k^m(z;\rho)\]
has non-empty compact values and is upper hemicontinuous. It can be shown that $(\rho, z)\mapsto \tilde \rho_k$ is continuous on $\text{Graph}(\Phi_k)$. 
Consequently, the composition $\Psi_k(\rho):=\tilde \rho_k(\rho, \Phi_k(\rho))$ is upper hemicontinuous with non-empty compact values. Finally, define $\Xi: \mathcal{P}([0,\infty])\rightarrow 2^{\mathcal{P}([0,\infty])}$ by
\[\Xi:=\sum_{k=1}^M p_k \Psi_k.\]
By \cite[Theorem 17.32]{InfDimAnalysis}, $\Xi$ is also upper hemicontinuous with non-empty compact values, and thus has closed graph. 
We are almost in a position to apply the Kakutani-Fan-Glicksberg fixed point theorem to $\Xi$ (see e.g.\ \cite[Corollary 17.55]{InfDimAnalysis}), except that we do not know if it is convex-valued or not. The model may need to be substantially modified to make this property hold. For this reason, we leave the heterogeneity at the inter-team level for future research.

\appendix

\section{Proofs}\label{sec:proof}

\subsection{Proof of Theorem~\ref{thm:EQMeffort}}
\begin{proof}
(i) The function $V_{\lambda, z}$ in \eqref{DPE2soln} is well-defined by the non-negativity and integrability of $\lambda$ (see the definition of $\cA$). It is straightforward to verify that $V_{\lambda, z}$ satisfies \eqref{DPE2} and equivalently, \eqref{DPE} with $\bar\alpha=\alpha_z$, and that $\alpha_z\in \cA$. Standard verification argument shows that $V_{\lambda, z}$ is the value function (in response to $(\lambda, z,\alpha_z)$), and that $\alpha_z$ is an optimal control.

(ii) Let $\hat \alpha \in \cA$ be any equilibrium control and $\hat V$ be the corresponding equilibrium value function within the team. Since the best response problem within a team is time-consistent, the restriction of $\hat\alpha$ on $[r,1]$ is optimal for $\hat V(r)$ (in response to $(\lambda, z,\hat \alpha)$) for any $r<1$.
By the optimality of $\hat \alpha$, we have $\hat V(r)\le G_z(r)$.
On the other hand, taking the admissible control $\alpha=\epsilon G_z\in\mathcal{A}$, we obtain
\[\hat V(r)\ge \beta G_z(1)-c \epsilon^2 \mathbb{E}\left[\int_{\rho(0)}^{\rho(\tau)} \frac{G^2_z(y)}{\lambda(y)(1-y)}dy\right]\ge  \beta G_z(1)-c \epsilon^2 \mathbb{E}\left[\int_{0}^{1} \frac{G^2_z(y)}{\lambda(y)(1-y)}dy\right].\]
Letting $\epsilon\rightarrow 0+$ yields $\hat V(r)\ge \beta G_z(1)$. Because $G_z(1-)=G_z(1)=0$, we must have $\hat V(1-)=0$.

Claim that $\hat V$ is absolutely continuous. Once this is proved, dynamic programming yields that $\hat V$ must a.e.\ satisfy \eqref{DPE} with $\bar\alpha=\hat\alpha$, and that $\hat \alpha$ coincides with $\alpha_z$, which further implies that $\hat V$ satisfies
\eqref{DPE2}. It is easy to check that \eqref{DPE2} has at most one absolutely continuous solution, namely, \eqref{DPE2soln}. By its uniqueness, we must have $\hat V=V_{\lambda, z}$.

The rest is devoted to the proof of absolutely continuity of $\hat V$ by a control-theoretical argument adapted from \cite{Nutz-Zhang.19}. Fix an arbitrary $r_0<1$. Since $\lambda$ is assumed to be locally piecewise Lipschitz and strictly positive on $[0,1)$, it is uniformly bounded away from zero on $[0,r_0]$. This implies that $\rho$ will reach $r_0$ in finite time. Let $0\le r<r+h\le r_0$, we wish to bound $\hat V(r)-\hat V(r+h)$ by a constant times $h$. There are two subtle differences from the proof in \cite{Nutz-Zhang.19}: First, due to the bonus payment, monotonicity of the value function is unclear; thus, a lower bound for $\hat V(r)-\hat V(r+h)$ is no longer trivial. Second, in our model a single member has negligible impact on the team's completion time. Hence $\tau^{r}$ and $\tau^{r+h}$ (where the superscript indicates the dependence on $\rho(0)$) are different regardless of how we choose a single member's control.

Denote by $\rho^r$ the state process starting at $\rho(0)=r$, and let $t_h$ be the first time $\rho^r$ hits $r+h$, which is finite. By the memoryless property of exponential random variables and the flow property of $\rho^r$, we have that
\begin{align*}
\P\left(\tau^r\ge t_h+s| \tau^r\ge t_h \right)&=\P\left(Z^i>\int_{0}^{t_h+s} z\hat \alpha (\rho^r(u))du\Big| Z^i>\int_{0}^{t_h} z\hat \alpha (\rho^r(u))du\right)\\
&=\P\left(Z^i>\int_{t_h}^{t_h+s} z\hat \alpha (\rho^r(u))du\right)=\P\left(Z^i>\int_{0}^{s} z\hat \alpha (\rho^r(t_h+u))du\right)\\
&=\P\left(Z^i>\int_{0}^{s} z\hat \alpha (\rho^{r+h}(u))du\right)=\P\left(\tau^{r+h}\ge s \right).
\end{align*}
In other words, the distribution of $\tau^r-t_h$ conditioned on the event $\tau^r\ge t_h$ is the same as the distribution of $\tau^{\tau+h}$. Using this, we deduce that
\begin{align*}
\hat V(r)&=J(r, \hat \alpha;\lambda, z, \hat \alpha)=\E\left[G_z(\rho^r(\tau^r))-c\int_0^{\tau^r}\hat\alpha(\rho^r(t))^2\,dt\, \right]\\
&\le \E\left[1_{\{\tau^r\le t_h\}} G_z(r)\right] + \P(\tau^r>t_h)\E\left[G_z(\rho^r(\tau^r))-c\int_{t_h}^{\tau^r}\hat \alpha(\rho^r(t))^2\,dt \Big| \tau^r>t_h\right]\\
&=\P(\tau^r\le t_h)G_z(r)+\P(\tau^r>t_h)\E\left[G_z(\rho^r(\tau^{r+h}+t_h))-c\int_{t_h}^{\tau^{r+h}+t_h}\hat \alpha(\rho^r(t))^2\,dt \right]\\
&=\P(\tau^r\le t_h)G_z(r)+\P(\tau^r>t_h)\E\left[G_z(\rho^{r+h}(\tau^{r+h}))-c\int_{0}^{\tau^{r+h}}\hat \alpha(\rho^{r+h}(t))^2\,dt \right]\\
&= \P(\tau^r\le t_h)G_z(r)+\P(\tau^r>t_h) J(r+h,\hat \alpha;\lambda, z, \hat \alpha) \\
&\le \P(\tau^r\le t_h)G_z(0)+\hat V(r+h).
\end{align*}
Similarly, with
$$
  \alpha := \begin{cases}
  \epsilon & \mbox{ on } [r,r+h),\\
  \hat {\alpha} & \mbox{ on } [r+h,1),
  \end{cases}
  $$
we can show that
\begin{align*}
\hat V(r)&\ge J(r,\alpha;\lambda, z, \hat \alpha)\\
&\ge \P(\tau^r> t_h) \E\left[G_z(\rho^r(\tau^r))\left(\beta+(1-\beta)\frac{\alpha}{\hat\alpha}(\rho^r(\tau^r))\right) -c\int_{t_h}^{\tau^r} \hat\alpha(\rho^r(t))^2\,dt \Big| \tau^r>t_h\right]- c  \epsilon^2 t_h\\
&=\P(\tau^r> t_h) \E\left[G_z(\rho^{r+h}(\tau^{r+h})) -c\int_{0}^{\tau^{r+h}}\hat \alpha(\rho^{r+h}(t))^2\,dt\right]- c  \epsilon^2 t_h\\
&=\P(\tau^r> t_h)J(r+h,\hat \alpha;\lambda, z, \hat\alpha)- c  \epsilon^2 t_h=\P(\tau^r> t_h)\hat V(r+h)- c  \epsilon^2 t_h\\
&\ge \hat V(t+h)-\P(\tau^r\le t_h)G_z(0)- c  \epsilon^2 t_h.
\end{align*}
Taking limit as $\epsilon\rightarrow 0+$ and combining the two chains of inequalities, we obtain
\[|\hat V(r)-\hat V(r+h)|\le \P(\tau^r\le t_h)G_z(0).\]
It remains to note that
\begin{align*}
\P(\tau^r\le t_h)&=1-\exp\left(-\int_0^{t_h}z\hat\alpha(\rho^r(s))ds\right)\\
&\le \int_0^{t_h}z\hat\alpha(\rho^r(s))ds=\int_r^{r+h} z\hat\alpha(y)d\rho^{-1}(y)=\int_r^{r+h} \frac{z\hat\alpha(y)}{\lambda(y)(1-y)}dy\\
&\le \esssup_{y\in[0, r_0]}\left|\frac{z\hat\alpha(y)}{\lambda(y)(1-y)}\right| h.
\end{align*}
We conclude that $\hat V$ is Lipschitz continuous on $[0, r_0]$ for any $r_0<1$ and thus, absolutely continuous on $[0,1)$.
\end{proof}

\subsection{Proof of Theorem~\ref{thm:manager}}

\begin{proof}
We only show (ii) and (iii). Suppose $K(1+p)\theta>\kappa_0$ so that zero is not an equilibrium team size. For each $\bar z >0$, define
\begin{equation}\label{eq:F}
\notag F_{\bar z}(z):=K(1+p)\theta \left[1+  p \left(\frac{z}{\bar z}\right)^{\eps-2} \right]^{-1}-k z^\delta, \quad z\ge 0.
\end{equation}
By \eqref{eq:z*0}, $z^\ast$ is an equilibrium team size if and only if $z^\ast\in \argmax_{z> 0}F_{z^\ast}(z)$ and $K\theta-\kappa_0-\kappa(z^\ast)\ge 0$.
Since $F_{\bar z}$ is continuous and $\lim_{z\rightarrow \infty}F_{\bar z}(z)=-\infty$, the maximum of $F_{\bar z}$ is attained either at $z=0$ or at some interior point where the first derivative
\[F'_{\bar z}(z)=\frac{K(1+p)\theta p(2-\eps)(z/\bar z)^{\eps-3}(\bar z)^{-1}}{\left(1+p(z/\bar z)^{\eps-2}\right)^2}- k\delta z^{\delta-1}\]
vanishes. Any positive equilibrium team size $z^\ast$ must satisfy
$F_{z^\ast}'(z^\ast)=0$, giving the unique candidate $z_m^\ast$ in \eqref{eq:z*m}. It remains to check that $F_{z_m^\ast}(z)$ attains global maximum at $z=z_m^\ast$ and that $K\theta-\kappa_0-\kappa(z_m^\ast)\ge 0$.

Let us rewrite the function
\[F_{z_m^\ast}(z)=K(1+p)\theta f(z/z_m^\ast),\]
where
\begin{equation*}\label{eq:f}
f(x):=\frac{x^{2-\eps}}{x^{2-\eps}+p}-\frac{(2-\eps)p}{\delta (1+p)^2}x^\delta, \quad x\ge 0.
\end{equation*}
$F_{z_m^\ast}(z)$ attains global maximum at $z=z_m^\ast$ if and only if $f(x)$ attains global maximum on $\R_+$ at $x=1$. We have
\[f'(x)=(2-\eps)px^{1-\eps}\left[\frac{1}{(x^{2-\eps}+p)^2}-\frac{x^{\eps+\delta-2}}{(1+p)^2}\right].\]
It is easy to see that for $x>0$, $\sgn(f'(x))=\sgn(h(x))$, where
\[h(x):=(1+p)x^{1-\frac{\eps+\delta}{2}}-x^{2-\eps}-p\]
Notice that $h(1)=0$ and $h'(1)=p+\eps-1-(1+p)(\eps+\delta)/2.$
Consider two cases:

(i) $\eps+\delta\ge 2$. In this case, $h$ is strictly decreasing, which implies $f'$ is positive when $0<x<1$ and negative when $x>1$. Consequently, the global maximum of $f$ is attained at $x=1$ as desired.

(ii) $\eps+\delta< 2$. In this case, $h$ is strictly concave, which implies that it can cross the $x$-axis at most twice. As $h(0)=-p<0$,  $x=1$ is a global maximum of $f$ if and only if $h'(1)<0$ and $f(1)\ge f(0)$, i.e.,
$$\delta>\frac{(2-\eps)p+\eps-2}{1+p}\quad\text{and}\quad\delta\ge \frac{(2-\eps)p}{(1+p)}.$$

Note that $\delta\ge (2-\eps)p/(1+p)$ is equivalent to $\eps+\delta\ge 2-\delta/p$. Combining the two cases, we see that $f(x)$ attains global maximum at $x=1$ if and only if $\delta\ge (2-\eps)p/(1+p)$. We also have that
\[K\theta-\kappa_0-\kappa(z_m^\ast)=K\theta-\kappa_0- \frac{K\theta p (2-\eps)}{\delta (1+p)}\ \ge 0\]
if and only if
\[\left[1-\frac{\kappa_0}{K\theta}\right]\delta \ge \frac{(2-\eps)p}{1+p},\]
which implies $\delta\ge (2-\eps)p/(1+p)$. The rest of the theorem statement follows from direct computation using \eqref{eq:alphaz2}, \eqref{eq:z*m}, \eqref{eq:Vm} and \eqref{eq:Vw}.
\end{proof}

\subsection{Proof of Theorem~\ref{thmpartner}}

\begin{proof}
\textbf{Part 1:} Let us first examine the candidate equilibrium team size $z^*=0$.

\textit{Case I}: $\epsilon=0$. For $\kappa_0=0$, the conclusion I(i) of Theorem~\ref{thmpartner} is easy to verify. Now assume $\kappa_0>0$ and $\delta\in(0,1]\cup[2,\infty)$. Let us define
\begin{align*}
J(z):=\frac{(1+\beta) K(1+p)}{2}-\frac{\kappa_0}{z}-kz^{\delta-1},\quad z>0.
\end{align*}
We get $J'(z)=\kappa_0z^{-2}-k(\delta-1)z^{\delta-2}$ and $J''(z)=-2\kappa_0z^{-3}-k(\delta-1)(\delta-2)z^{\delta-3}<0$ as $\delta\leq 1$ or $\delta\geq 2$. Therefore, the unique interior critical point $\hat{z}:=\left(\frac{\kappa_0}{k(\delta-1)}\right)^{\frac{1}{\delta}}$ is the global maximum point. We have that
\begin{align*}
\hat{z}J(\hat{z})=\frac{(1+\beta) K(1+p)}{2}\left(\frac{\kappa_0}{k(\delta-1)}\right)^{\frac{1}{\delta}} -\kappa_0-k\left(\frac{\kappa_0}{k(\delta-1)}\right).
\end{align*}
Then $z^*=0$ is the equilibrium team size if and only if $\hat z J(\hat z)\leq 0$, which is equivalent to
$$\frac{2^\delta(\kappa_0)^{\delta-1}k\delta^{\delta}}{(1+\beta)^{\delta}K^{\delta}(1+p)^{\delta}(\delta-1)^{\delta-1}}\geq 1.$$

\textit{Case II}: $\epsilon=1$. It is clear that $z^*=0$ is an equilibrium team size if $\frac{(1+\beta)}{2} K(1+p)\leq \kappa_0$. Now suppose that $\frac{(1+\beta)}{2} K(1+p)>\kappa_0$ and let us define
\begin{align*}
J(z):=\frac{\frac{(1+\beta)}{2} K(1+p)-\kappa_0}{z}-kz^{\delta-1},\quad z>0.
\end{align*}
We get
\begin{align*}
\lim_{z\rightarrow 0+}\frac{J(z)}{1/z}=\lim_{z\rightarrow 0+}zJ(z)=\lim_{z\rightarrow 0+} \frac{(1+\beta)}{2} K(1+p)-\kappa_0-kz^{\delta}=\frac{(1+\beta)}{2} K(1+p)-\kappa_0.
\end{align*}
It then follows that $\lim_{z\rightarrow 0+}J(z)=+\infty$ and hence $z^*=0$ is not an equilibrium team size in view of its definition.

\textbf{Part 2:} Next, consider the candidate equilibrium team size $z^*>0$. We can compute from \eqref{partnerH} that
\begin{align*}
H'(z;z^*)=&\frac{(1-\varepsilon)K(1+p)(1+\beta)}{(z^*)^{2-\varepsilon}} \frac{z^{1-2\varepsilon}}{p+(\frac z{z^*})^{2-\varepsilon}}- \frac{(2-\varepsilon)K(1+p)(1+\beta)}{2(z^*)^{4-2\varepsilon}} \frac{z^{3-3\varepsilon}}{(p+(\frac z{z^*})^{2-\varepsilon})^2} \notag\\
&+\frac{\kappa_0}{z^2}+k(1-\delta)z^{\delta-2}.
\end{align*}

Again as team sizes are required to be positive, we only need to consider interior maxima of $H(z;z^*)$. Therefore, $z^*$ is the equilibrium team size with partnership implies that $z^*$ satisfies that $H'(z^*;z^*)=0$, which gives that $z^*$ solves the algebraic equation
\begin{align}\label{algebraeq}
\left[ (1-\varepsilon)K(1+\beta)-\frac{(2-\varepsilon)K(1+\beta)}{2(p+1)} \right](z^*)^{1-\varepsilon}+k(1-\delta)(z^*)^{\delta}+\kappa_0=0.
\end{align}

\textit{Case I}: $\varepsilon=0$.
Suppose that $\delta\geq 3$. If $\kappa_0=0$, it is clear that the algebraic equation \eqref{algebraeq} admits a unique positive root $z^*_p=\left(\frac{A}{k(\delta-1)}\right)^{\frac{1}{\delta-1}}$, where we denote $A:=\frac{pK(1+\beta)}{(p+1)}>0$. If $\kappa_0>0$, let us denote $\gamma(x):=Ax+k(1-\delta)x^{\delta}+\kappa_0$. We have that $\lim_{x\rightarrow 0} \gamma(x)=\kappa_0>0$ and $\lim_{x\rightarrow \infty} \gamma(x)=-\infty$. Therefore, the equation $\gamma(x)=0$ admits at least one positive root. Moreover, we also know that $\gamma'(x)=A+k\delta(1-\delta)x^{\delta-1}$ and therefore $\gamma(x)$ is strictly increasing for $x\leq x^*$ and strictly decreasing for $x>x^*$, where
\begin{align}\label{xstar}
x^*:=\left(\frac{A}{k\delta(\delta-1)}\right)^{\frac{1}{\delta-1}}.
\end{align}
It then follows that the curve $y=\gamma(x)$ only hits $x$-axis once, which implies that $\gamma(x)=0$ admits a unique positive root $z_p^*$.

\textit{Case II}: $\varepsilon=1$. The algebraic equation \eqref{algebraeq} can be simplified as
\begin{align*}
-\frac{K(1+\beta)}{2(p+1)}+k(1-\delta)(z^*)^{\delta}+\kappa_0=0.
\end{align*}
It is clear that if 
$\delta\geq 2$ and $\frac{K(1+\beta)}{2(p+1)}<\kappa_0$, we can obtain the unique positive solution given in \eqref{equisol-size}.

It then suffices to verify that $H(z;z^*_p)$ attains its global maximum at the unique point $z=z^\ast_p$ in two cases.

\textit{Case I}: $\varepsilon=0$. Let us assume that $\delta\geq 3$ and $p\geq 1/3$. We first have
\begin{align*}
H(z;z_p^*)=\frac{K(1+p)(1+\beta)}{2} \frac{(\frac z{z^*_p})^{2}}{p+(\frac z{z^*_p})^{2}}-\frac{\kappa_0+kz^\delta}{z},
\end{align*}
and
\begin{align*}
H'(z;z_p^*)=&\frac{K(1+p)(1+\beta)}{(z_p^*)^2} \frac{z}{p+(\frac z{z_p^*})^{2}}- \frac{K(1+p)(1+\beta)}{(z_p^*)^{4}} \frac{z^{3}}{(p+(\frac z{z_p^*})^{2})^2} \notag\\
&+\frac{\kappa_0}{z^2}+k(1-\delta)z^{\delta-2}.
\end{align*}

It is straightforward to verify that the sign of $H'(z;z_p^*)$ coincides with the sign of $h(z;z_p^*)$, which is defined by
\begin{align*}
h(z;z_p^*)&:=pK(1+p)(1+\beta)\left(\frac z{z_p^*}\right)^{3}+ \left(\frac z{z_p^*}\right)\left(\frac{\kappa_0}{z}+k(1-\delta)z^{\delta-1}\right) \left(p+\left(\frac z{z_p^*}\right)^{2}\right)^2\\
=& pK(1+p)(1+\beta)\left(\frac z{z_p^*}\right)^{3}+ \left(\frac z{z_p^*}\right)\left(\frac{\kappa_0}{\frac{z}{z_p^*}}\frac{1}{z_p^*}+k(1-\delta)\left(\frac{z}{z_p^*}\right)^{\delta-1}(z_p^*)^{\delta-1}\right) \left(p+\left(\frac z{z_p^*}\right)^{2}\right)^2.
\end{align*}
Note that $z_p^*$ solves the equation $Ax+k(1-\delta)x^{\delta}+\kappa_0=0$, we get that
\begin{align*}
h(z;z_p^*)=pK(1+p)(1+\beta)\left(\frac z{z_p^*}\right)^{3}+ \left[-A\left(\frac{z}{z_p^*}\right)^{\delta}+\left(1-\left(\frac{z}{z_p^*}\right)^{\delta} \right)\kappa_0(z_p^*)^{-1} \right] \left(p+\left(\frac z{z_p^*}\right)^{2}\right)^2.
\end{align*}
After changing variable $x=\frac z{z_p^*}$, we can consider the function
\begin{align*}
h(x)&=pK(1+p)(1+\beta)x^{3}+ \left[-Ax^{\delta}+\left(1-x^{\delta} \right)\kappa_0(z_p^*)^{-1} \right] \left(p+x^{2}\right)^2\\
&=Bx^{3}+ \left[ C -(A+C ) x^{\delta} \right] \left(p+x^{2}\right)^2,
\end{align*}
with $B:=pK(1+p)(1+\beta)$ and $C:=\kappa_0(z_p^*)^{-1}$. First, we have $h(1)=B-(p+1)^2A=0$ by recalling that $A=\frac{pK(1+\beta)}{(p+1)}$. Moreover, we have that
\begin{align*}
h'(1)=3B-(p+1)^2(A+C)\delta -4A(p+1).
\end{align*}
As $A,C>0$ and $\delta\geq 3$, it follows that $(A+C)\delta\geq 3A$. As $4A(p+1)>0$, we can then deduce that $h'(1)<3B-3A(p+1)^2=0$ and hence $z=z_p^*$ is a local maximum of the function $H(z;z_p^*)$.

We then claim that the equation $h(x)=0$, $x>0$, admits a unique solution at $x=1$. As we already know that $h(1)=0$ and $h'(1)<0$, we will show that for any other $\bar{x}>0$ such that $h(\bar{x})=0$, we always have $h'(\bar{x})<0$ and therefore $\bar{x}=1$ must be the unique solution as $h(x)$ is a continuous function. Let us then assume that $\bar{x}\neq 1$ that also satisfies
\begin{align*}
B\bar{x}^{3}+ \left[ C -(A+C ) \bar{x}^{\delta} \right] \left(p+\bar{x}^{2}\right)^2=0,
\end{align*}
and we have
\begin{align*}
\frac{\bar{x}}{p+\bar{x}^2}h'(\bar{x})&=\left[3B\bar{x}^3-(A+C)\delta \bar{x}^{\delta}(p+\bar{x}^2)^2+[C-(A+C)\bar{x}^{\delta}]4\bar{x}^2(p+\bar{x}^2)\right]\frac{1}{p+\bar{x}^2}\\
&=-3 \left[ C -(A+C ) \bar{x}^{\delta} \right] \left(p+\bar{x}^{2}\right)-(A+C)\delta \bar{x}^{\delta}(p+\bar{x}^2)+[C-(A+C)\bar{x}^{\delta}]4\bar{x}^2\\
&=C\left[\bar{x}^2-3p- (1+\delta)\bar{x}^{2+\delta}+(3-\delta)p\bar{x}^{\delta}\right]+A\left[(-1-\delta)\bar{x}^{2+\delta}+(3-\delta)p\bar{x}^{\delta}\right]\\
&=C\left[(\bar{x}^2-3p)(1-\bar{x}^{\delta})-\delta \bar{x}^{\delta}(\bar{x}^2+p) \right] +A\left[-\bar{x}^{\delta}\left((1+\delta)\bar{x}^2+(\delta-3)p\right)\right]=:g(\bar{x}).
\end{align*}
To show $h'(\bar{x})<0$ for $\bar{x}\neq 1$, it is equivalent to show that $g(\bar{x})<0$. First, for the second term of $g(\bar{x})$, the condition $\delta\geq 3$ clearly implies that $(1+\delta)\bar{x}^2+(\delta-3)p>0$ for any $x>0$ and therefore the second term is always negative for $x>0$. For the first term of $g(\bar{x})$, recall the condition that $p\geq 1/3$ and hence $\sqrt{3p}\geq 1$. It is then clear to see that either if $\bar{x}\leq 1$ or $\bar{x}\geq \sqrt{3p}$, we have $(\bar{x}^2-3p)(1-\bar{x}^{\delta})\leq 0$ and it follows that the first term is nonpositive. Otherwise, for $1<\bar{x}<\sqrt{3p}$, we can also write
$$\bar{x}^2-3p- (1+\delta)\bar{x}^{2+\delta}+(3-\delta)p\bar{x}^{\delta}\leq\bar{x}^2-3p- (1+\delta)\bar{x}^{2}+(3-\delta)p<0.$$

This verifies the claim that $g(\bar{x})<0$ for any $\bar{x}\geq 0$. We can then conclude that the claim holds and $x=1$ is the unique solution such that $h(x)=0$. It follows that $H(z;z_p^*)$ admits a unique critical point $z=z_p^*$ and hence the local maximum is also the global maximum.

\textit{Case II}: $\varepsilon=1$. We have that
\begin{align*}
H(z;z_p^*)=\frac{K(1+p)(1+\beta)}{2}\frac{1}{pz_p^*+z}-\frac{\kappa_0+kz^\delta}{z},
\end{align*}
as well as
\begin{align*}
H'(z;z_p^*)=\frac{-K(1+p)(1+\beta)}{2}\frac{1}{(pz_p^*+z)^2}+\frac{\kappa_0}{z^2}+k(1-\delta)z^{\delta-2}.
\end{align*}
It is clear that the sign of the function $H'(z;z_p^*)$ coincides with the sign of the function
\begin{align*}
h(z;z_p^*):=\frac{-K(1+p)(1+\beta)}{2}\left(\frac{z}{z_p^*}\right)^2+\left[\kappa_0+k(1-\delta)(z_p^*)^{\delta} \left(\frac{z}{z_p^*}\right)^{\delta}\right] \left[p+\frac{z}{z_p^*}\right]^2.
\end{align*}
After changing variable, we can consider the function
\begin{align*}
h(x)=A x^2+B\left(x^{2+\delta}+2x^{1+\delta}+x^{\delta} \right)+\kappa_0(p+x)^2,
\end{align*}
where $A:=\frac{-K(1+p)(1+\beta)}{2}$ and $B:=(z_p^*)^{\delta}k(1-\delta)=\frac{K(1+\beta)}{2(p+1)}-\kappa_0$.

Note that $h(1)=0$. Moreover, we have that
\begin{align*}
h'(1)=\left[\frac{2(1+\delta)}{(p+1)^2}-1\right]K(1+p)(1+\beta)-\kappa_0\left[4(1+\delta)-2(p+1)\right].
\end{align*}
By the assumption that $2(1+\delta)>(p+1)^2$, we get $2(1+\delta)>(p+1)$ as well and hence $\frac{\frac{2(1+\delta)}{(p+1)^2}-1}{4(1+\delta)-2(p+1)}\leq \frac{1}{2(p+1)^2}$. We obtain that $h'(1)<\left[ \frac{K(1+\beta)}{2(p+1)}-\kappa_0\right] \left[4(1+\delta)-2(p+1)\right]<0$ as we assume $\frac{K(1+\beta)}{2(p+1)}<\kappa_0$. Again, it follows that $z_p^*$ is a local maximum of the function $H(z;z_p^*)$.

We then claim that the equation $h(x)=0$, $x>0$, admits a unique solution. We again show that for any point $\bar{x}$ such that $h(\bar{x})=0$, we always have $h'(\bar{x})<0$. Let us assume that $\bar{x}$ satisfies
\begin{align*}
A \bar{x}^2+B\left(\bar{x}^{2+\delta}+2\bar{x}^{1+\delta}+\bar{x}^{\delta} \right)+\kappa_0(p+\bar{x})^2=0,
\end{align*}
and check that
\begin{align*}
\bar{x}h'(\bar{x})=&2A\bar{x}^2+B\left((2+\delta)\bar{x}^{2+\delta}+2(1+\delta)\bar{x}^{1+\delta}+\delta \bar{x}^{\delta} \right)+\kappa_02\bar{x}(p+\bar{x})\\
=&-2B\left(\bar{x}^{2+\delta}+2\bar{x}^{1+\delta}+\bar{x}^{\delta} \right)-2\kappa_0(p+\bar{x})^2+B\left((2+\delta)\bar{x}^{2+\delta}+2(1+\delta)\bar{x}^{1+\delta}+\delta \bar{x}^{\delta} \right)\\
&+\kappa_02\bar{x}(p+\bar{x})\\
=&B\left( \delta\bar{x}^{2+\delta}+(2\delta-2)\bar{x}^{1+\delta}+(\delta-2) \bar{x}^{\delta}\right)-2p\kappa_0(p+\bar{x}).
\end{align*}
As $\delta\geq 2$, the quadratic function $\delta\bar{x}^{2}+(2\delta-2)\bar{x}+(\delta-2)>0$ for any $x>0$. Thanks to $B<0$, we have $h'(\bar{x})<0$ for any $\bar{x}>0$ if $h(\bar{x})=0$. This leads to the fact that $\bar{x}=1$ is the unique solution to the equation $h(x)=1$ as $h(x)$ is a continuous function. It then yields that the function $H(z;z_p^*)$ admits a unique critical point. Therefore, $z_p^*$ is the global maximum of $H(z;z_p^*)$, which completes the proof.
\end{proof}

\subsection{Proof of Proposition \ref{p1}}

\begin{proof}
Recall the function $H$ defined in \eqref{partnerH}. $z_p^*>0$ is an equilibrium only if $H'(z;z_p^*)|_{z=z_p^*}=0$, which is equivalent to
\begin{equation}\label{e151}
K\alpha(\eps,p)(1+\beta)(z_p^*)^{1-\eps}=-\kappa_0+k(\delta-1)(z_p^*)^\delta,
\end{equation}
where
$$\alpha(\eps,p):=1-\eps-\frac{2-\eps}{2(p+1)}.$$
As $\delta>1$, the functions $z\mapsto K\alpha(\eps,p)(1+\beta)z^{1-\eps}$ and $z\mapsto -\kappa_0+k(\delta-1)z^\delta$ have at most one intersection point $z_p^*>0$ for $z>0$.

Note that $\alpha(\eps,p)>0 \Leftrightarrow 2p-2p\eps-\eps>0$, and if this holds, it can be easily seen that $z_p^*$ is increasing w.r.t. $\beta$. Similarly we can show the monotonicity of $z_p^*$ w.r.t. $\beta$ when $\alpha(\eps,p)=0$ and $\alpha(\eps,p)<0$.
\end{proof}

\subsection{Proof of Proposition \ref{p2}}

\begin{proof}
Consider
\begin{align*}
\frac{d}{d\beta}V^p(\beta,z_*^p(\beta))&=\frac{\partial}{\partial\beta}V^p(\beta,z_*^p(\beta))+\frac{\partial}{\partial z}V^p(\beta,z_*^p(\beta))\cdot\frac{d}{d\beta}z_p^*(\beta)\\
&=\frac{K}{2}(z_p^*)^{-\eps}+\frac{\partial}{\partial z}V^p(\beta,z_*^p(\beta))\cdot\frac{d}{d\beta}z_p^*(\beta).
\end{align*}
We have that
\begin{align*}
\frac{\partial}{\partial z}V^p(\beta,z_*^p(\beta))&=(z_p^*)^{-2}\left[-\frac{K(1+\beta)\eps}{2}(z_p^*)^{1-\eps}+\kappa_0-k(\delta-1)(z_p^*)^\delta\right]\\
&=-(z_p^*)^{-2}\left[\frac{K(1+\beta)\eps}{2}(z_p^*)^{1-\eps}+K\alpha(\eps,p)(1+\beta)(z_p^*)^{1-\eps}\right]\\
&=-(z_p^*)^{-1-\eps}\cdot\frac{K(1+\beta)}{2(p+1)}\cdot p(2-\eps),
\end{align*}
where the second equality follows from \eqref{e151}. Therefore,
$$\frac{d}{d\beta}V^p(\beta,z_*^p(\beta))=\frac{K(z_p^*)^{-1-\eps}}{2}\left[z_p^*-\frac{1+\beta}{1+p}\cdot p(2-\eps)\cdot\frac{d}{d\beta}z_p^*(\beta)\right].$$
By Proposition \ref{p1}, when $2p-2p\eps-\eps\leq 0$, $\frac{d}{d\beta}z_p^*(\beta)\leq 0$ and thus $\frac{d}{d\beta}V^p(\beta,z_*^p(\beta))\geq 0$. For the rest of the proof, we assume $2p-2p\eps-\eps>0$.

By \eqref{e151} we have that
$$\frac{d}{d\beta}z_p^*(\beta)=\frac{K\alpha(\eps,p)}{\kappa_0(1-\eps)(z_p^*)^{\eps-2}+k(\delta-1)(\delta+\eps-1)(z_p^*)^{\delta+\eps-2}}.$$
Since $2p-2p\eps-\eps>0\Leftrightarrow \alpha(\eps,p)> 0$,
we deduce that
$$\frac{d}{d\beta}z_p^*(\beta)\leq\frac{K\alpha(\eps,p)}{k(\delta-1)(\delta+\eps-1)(z_p^*)^{\delta+\eps-2}}.$$
It follows that
\begin{align*}
\frac{d}{d\beta}V^p(\beta,z_*^p(\beta))&\geq\frac{K(z_p^*)^{-1-\eps}}{2}\left[z_p^*-\frac{1+\beta}{1+p}\cdot p(2-\eps)\cdot\frac{K\alpha(\eps,p)}{k(\delta-1)(\delta+\eps-1)(z_p^*)^{\delta+\eps-2}}\right]\\
&=\frac{K(z_p^*)^{-1-\eps}}{2}\left[z_p^*-\frac{p(2-\eps)}{1+p}\cdot\frac{-\kappa_0(z_p^*)^{\eps-1}+k(\delta-1)(z_p^*)^{\delta+\eps-1}}{k(\delta-1)(\delta+\eps-1)(z_p^*)^{\delta+\eps-2}}\right]\\
&\geq\frac{K(z_p^*)^{-\eps}}{2}\left[1-\frac{p(2-\eps)}{(1+p)(\delta+\eps-1)}\right],
\end{align*}
where the second equality follows from \eqref{e151}. This completes the proof.
\end{proof}

\begin{acknowledgements}
  Yuchong Zhang is supported by NSERC Discovery Grant RGPIN-2020-06290.
\end{acknowledgements}

%
%

\bibliography{team}
\bibliographystyle{spmpsci}      


\end{document}